\documentclass[3p,times,12pt]{elsarticle}
\usepackage[font=normalsize,labelfont=bf]{caption}

\usepackage{hyperref}

\journal{Computers \& Fluids}

\biboptions{sort&compress} 

\usepackage{amssymb,amsmath}
\usepackage{bm}

\usepackage{textcomp}
\usepackage{gensymb}

\usepackage{subcaption}

\usepackage{longtable,booktabs,tabularx,tabulary}
\newcolumntype{L}[1]{>{\hsize=#1\hsize\raggedright\arraybackslash}X}%
\newcolumntype{R}[1]{>{\hsize=#1\hsize\raggedleft\arraybackslash}X}%
\newcolumntype{C}[2]{>{\hsize=#1\hsize\columncolor{#2}\centering\arraybackslash}X}%
\newcolumntype{Y}{>{\centering\arraybackslash}X}

\usepackage[status=draft,index,multiuser]{fixme}
\fxsetface{margin}{\linespread{1}\scriptsize}
\fxusetheme{color}
\FXRegisterAuthor{bz}{abz}{BZ}  
\FXRegisterAuthor{aw}{aaw}{AW}  
\FXRegisterAuthor{km}{akm}{KM}  
\FXRegisterAuthor{gb}{agb}{GB}  
\usepackage{soul}

\begin{document}

\begin{frontmatter}

\title{Immersed boundary finite element method for blood flow simulation}


\address[uwa]{Intelligent Systems for Medicine Laboratory,
The University of Western Australia,\\
35 Stirling Highway, Perth, Western Australia}

\address[patras]{Department of Physics, University of Patras,
Patras, 26500, Rion, Greece}

\address[harvard]{Harvard Medical School, Boston, Massachusetts,
USA}

\author[uwa]{G.C. Bourantas\corref{mycorrespondingauthor}}
\ead{george.bourantas@uwa.edu.au}

\author[patras]{D.L. Lampropoulos}

\author[uwa]{B.F. Zwick}

\author[patras]{V.C. Loukopoulos}

\author[uwa]{A. Wittek} 

\author[uwa,harvard]{K. Miller}

\cortext[mycorrespondingauthor]{Corresponding author}

\begin{abstract}
We present an efficient and accurate immersed
boundary (IB) finite element (FE) solver for numerically solving
incompressible Navier--Stokes equations. Particular emphasis is given to
internal flows with complex geometries (blood flow in the vasculature
system). IB methods are computationally costly for internal flows,
mainly due to the large percentage of grid points that lie outside the
flow domain. In this study, we apply a local refinement strategy, along
with a domain reduction approach in order to reduce the grid that
covers the flow domain and increase the percentage of the grid nodes
that fall inside the flow domain. The proposed method utilizes an
efficient and accurate FE solver with the incremental
pressure correction scheme (IPCS), along with the boundary condition
enforced IB method to numerically solve the transient,
incompressible Navier--Stokes flow equations. We verify the accuracy of
the numerical method using the analytical solution for Poiseuille flow
in a cylinder. We further examine the
accuracy and applicability of the proposed method by considering flow within complex geometries, such as blood flow in aneurysmal vessels
and the aorta, flow configurations which would otherwise be 
extremely difficult to solve by most IB methods. Our method
offers high accuracy, as demonstrated by the verification examples, and high efficiency, as demonstrated through the solution of blood flow within complex geometry on an off-the-shelf laptop computer.
\end{abstract}


\end{frontmatter}



\section{Introduction}

Immersed boundary methods have been used to simulate flow past rigid and moving/deforming bodies with irregular/complex shapes using Cartesian grids. IB methods, in contrast to mesh-based methods,
avoid tedious mesh generation, since they rely on Cartesian grids to
solve the governing flow equations, and on discrete set of points (following the literature \cite{peskin_1972_flow}, referred to them as Lagrangian points) for the imposition of prescribed velocity boundary condition. The IB
methods are particularly efficient for moving objects and deforming
geometries --- they are attractive with moving boundaries as they avoid
re-meshing --- where it does not require the generation a new mesh at each
time step, but only the updated position of the points describing the
immersed boundary. However, high computational cost of most currently used IB methods in application to internal flows remains a challenge.
\citep{anupindi_etal_2013_novel,zelicourt_etal_2009_flow,zhu_etal_2019_graphpartitioned}.

\subsection{State-of-the art of the existing IB methods}

In IB methods, flow equations are discretized on a Cartesian grid
that does not conform with the boundaries of the immersed object,
and the boundary conditions are imposed indirectly through modifications of
the governing Navier--Stokes (N-S) equations.
The modification applies as a forcing function in the governing equations
that incorporates the presence of the immersed boundary. According to a
forcing function \(\bm{f}\left( \bm{x,}t \right)\) formulation, the existing
IB methods
\citep{mittal_iaccarino_2005_immersed}
are divided in two groups: 
\emph{continuous (or feedback) forcing} and 
\emph{discrete (or direct)} \emph{forcing} methods.
These two approaches are often called
diffuse and sharp interface methods,
respectively.

\subsubsection{Continuous forcing approach}

In the continuous forcing approach, the forcing function
\(\bm{f}\left(\bm{x,}t \right)\) is included into the momentum
equation (N-S equations). This approach was introduced by
\citet{peskin_1972_flow,peskin_1977_numerical}
to model blood flow through a beating heart. Peskin's method
is a mixed Eulerian--Lagrangian finite difference method for computing
the flow interaction with a flexible immersed boundary. The fluid flow
is governed by the incompressible N-S equations, numerically solved on a
stationary Cartesian grid, while the immersed boundary is represented by
a set of massless elastic fibres. The locations of the fibres is tracked
in a Lagrangian fashion by a collection of massless points that move
with the local fluid velocity.

Peskin's method was later applied to rigid bodies by
\citet{goldstein_etal_1993_modeling}.
In this method, often called virtual boundary method, the
immersed boundary is treated as a fictitious boundary embedded into the
fluid. The embedded boundary applies force to the fluid, such that the
fluid will be at rest on the surface (no-slip condition). The force
\(\bm{F}\left( s,t \right)\) on the boundary is computed such that
the fluid velocity \(\bm{u}\left( \bm{x},t \right)\) satisfies
the no-slip condition on the boundary. The body force, is not known
\emph{a priori} and is computed through the velocity on the immersed
boundary in a feedback forcing manner.

\citet{saiki_biringen_1996_numerical}
extended the feedback forcing approach, and
eliminated the spurious oscillations caused by the applied feedback
forcing term at the boundary. The force \(\bm{F}\left( s,t \right)\)
on the boundary has been modified, and a more accurate interpolation of
the fluid velocity at the boundary points has been developed. A similar
approach to the virtual boundary method has been introduced by
\citet{lai_peskin_2000_immersed}.
In this approach, to simulate the flow around a
rigid boundary, the boundary is not fixed and undergoes small movement. The
immersed boundary points \(\bm{X}\) are connected to fixed
equilibrium points \(\bm{X}^{\text{eq}}\) using a very stiff spring
with large stiffness constant. In case the boundary points move away
from the desired location, the spring force will pull these boundary
points back.

\citet{khadra_etal_2000_fictitious}
developed the penalty continuous forcing approach. The main idea, is
that the entire flow occurs in a porous medium, governed by the
Navier--Stokes--Brinkman equations. The equations contain an additional
term of volume drag, called Darcy drag, which accounts for the action of
the porous medium on the flow.
\citet{su_etal_2007_immersed}
proposed a new
implicit force formulation on the Lagrangian marker to ensure exact
imposition of no-slip boundary condition at the immersed boundary. The
immersed solid boundary is represented by discrete Lagrangian markers,
which apply forces to the Eulerian fluid domain. The Lagrangian markers
and the fluid variables on the fixed Eulerian grid exchange information
through a simple discrete delta function.

\citet{glowinski_etal_1998_distributed}
proposed the distributed Lagrange
multiplier method (DLM). The method uses a finite element (FE) method
(projection or predictor-corrector method) and introduces Lagrange
multipliers (i.e. body force) on the immersed boundary to satisfy the
no-slip condition.
\citet{lee_leveque_2003_immersed}
developed the immersed
interface method (IIM), a method that was initially developed for
elastic membranes. In the IIM, the boundary force/force strength
\(\bm{F}\left( \bm{s},t \right)\) is decomposed into tangential
and normal components. The interface is explicitly tracked in a
Lagrangian manner. The tangential component of the force is included in
the momentum equation as an explicit term and the explicit normal
boundary force is implemented into the governing equations in terms of a
pressure jump condition across the interface
\citep{taira_colonius_2007_immersed}.

\subsubsection{Discrete forcing approach}

In the discrete forcing approach, the flow equations are discretized on a
Cartesian grid neglecting the immersed boundary. Instead, the discretization at the Cartesian grid points near the immersed boundary is adjusted to account for the presence of the immersed boundary. This discrete approach is is better suited for higher Reynolds numbers, since
the velocity boundary conditions at the immersed boundary are imposed
without introducing any forcing term.

\citet{mohd-yusof_1997_combined}
developed a spectral method, which uses a forcing
term that is numerically computed using the difference between the
interpolated and the prescribed velocity on the immersed boundary
points. This way, the errors between the calculated and the prescribed
velocity on the immersed boundary are compensated by the forcing term.
\citet{fadlun_etal_2000_combined}
extended the discrete-time forcing
approach introduced by
\citet{mohd-yusof_1997_combined}
to a three-dimensional
finite difference method on a standard marker-and-cell (MAC) staggered
grid and showed that the approach was more efficient than
feedback-forcing.
\citet{balaras_2004_modeling}
proposed a better reconstruction
scheme, based on the method of
\citet{mohd-yusof_1997_combined}
and
\citet{fadlun_etal_2000_combined},
which applies spreading/interpolation along the line
normal to the immersed boundary. The algorithm eliminates the
ambiguities associated with interpolation along the Cartesian grid
lines, but it is limited only to flows with immersed boundaries aligned
with one coordinate direction.
\citet{gilmanov_etal_2003_general}
developed
a new reconstruction scheme, which applies to complex, three-dimensional
immersed boundaries. The proposed scheme maintains a sharp
fluid/immersed boundary interface by discretizing the immersed boundary
using an triangular mesh. The numerical solution on the Cartesian grid
points close to the immersed boundary points (Lagrangian points) is
reconstructed via linear interpolation along the boundary surface normal
vector.

\citet{zhang_zheng_2007_improved}
developed an improved version of
Mohd-Yusof's method by developing a bilinear interpolation/extrapolation
function to interpolate force. Herein, both tangential and the normal
velocity components (as opposed to Mohd-Yusof's method) are used to
interpolate force. Therefore, the accuracy in the vicinity of the
immersed boundary is enhanced without using higher-order schemes or
using body-fitted grids near the immersed boundary. 
\citet{choi_etal_2007_immersed}
developed a finite-volume immersed boundary method valid to all
Reynolds numbers and is suitable for implementation on arbitrary grid
topologies. They introduced the concept of tangency correction by
decomposing the velocity into tangential and normal components along the
outward normal direction to the immersed surface. The tangential
velocity component is expressed as a power-law function (which is rather
arbitrary) of the wall normal distance.

\subsection{Contributions of this study}

In the present study we combine the boundary condition-enforced immersed
boundary (BCE-IB) method
\citep{ghommem_etal_2020_hydrodynamic,wu_shu_2009_implicit}
with the incremental pressure correction
scheme (IPCS)
\citep{goda_1979_multistep}
using the finite element (FE) method. The main advantage of the BCE-IB method is that it satisfies accurately both the the governing equations and boundary conditions using velocity and pressure correction procedures. The velocity correction applies implicitly such that the velocity on the immersed boundary (Lagrangian points) interpolated from the corrected velocity values computed on the mesh nodes (Eulerian nodes) accurately satisﬁes the prescribed velocity boundary conditions. The incremental pressure correction scheme (IPCS) is a modified version of projection method, which provides improved accuracy at little extra computational cost.  We implement the method using FEniCS
\citep{unknown_2019_fenics}
(an open-source software package that solves partial
differential equations) to numerically solve the incompressible Navier-Stokes
equations. The proposed IB scheme applies to both external and internal
flows, but in the present study we are particularly interested in
internal flow problems.

Our numerical scheme, increases the computational efficiency of the IB
method for internal flow cases, by reducing the number of unused elements
through a sophisticated and efficient mesh refinement inside the immersed boundary. Reducing the computational cost
is of paramount importance for IB methods applied to internal flows,
since the advantageous features of IB methods, that have made them popular for external flow simulations, do not entirely apply for
internal flows. Users of IB methods are aware of this problem and have
come up with various strategies to circumvent this liability
\citep{anupindi_etal_2013_novel,zelicourt_etal_2009_flow,zhu_etal_2019_graphpartitioned}.

The remainder of the paper is organized as follows.
In Section~2,
we present the numerical formulation for the proposed immersed boundary (IB) method.
In Section 3,
we solve benchmark problems to verify the accuracy of the proposed IB method.
Section 4 is concerned with the numerical
examples that demonstrate the accuracy and applicability of the proposed
IB method. Section 5 contains conclusions.

\section{Numerical formulation of the proposed immersed boundary method}

\subsection{Governing equations}

We consider the incompressible viscous flow in a spatial domain
\(\Omega\), which contains an immersed boundary in the form of a closed
surface \(\partial S\). The immersed boundary is modeled as localized
body forces acting on the surrounding fluid. The incompressible N-S
equations --- in their primitive variables (velocity \(\bm{u}\) and
pressure \(p\)) formulation --- accounting for the immersed boundary are
written as:
\begin{equation}
    \label{eq:momentum}
    \rho \left(
        \frac{\partial\bm{u}}{\partial t}
        + \bm{u} \cdot \bm{\nabla}\bm{u}
    \right) = 
    - \nabla p 
    + \nabla \cdot 2v\bm{\varepsilon}(\bm{u}) 
    + \bm{f},
\end{equation}
\begin{equation}
    \label{eq:incompressibility-constraint}
    \nabla \cdot \bm{u} = 0,
\end{equation}
subject to the no-slip boundary condition on \(\partial S\)
\begin{equation}
    \label{eq:noslip_bc}
    \bm{u}(\bm{X}(s),t) = \bm{U}_{B},
\end{equation}
with \(\bm{U}_{B}\) being the prescribed velocity of the immersed boundary, and
\(v\) the kinematic viscosity of the fluid. The term
\(\bm{\varepsilon}(\bm{u})\) is the strain-rate tensor
defined as:
\begin{equation}
    \bm{\varepsilon}(\bm{u}) = \frac{1}{2}\left(
        \nabla\bm{u} + \left( \nabla\bm{u} \right)^\mathrm{T}
    \right).
\end{equation}

The forcing term \(\bm{f}\) is added to the right hand side (RHS) of
the momentum Eq.\eqref{eq:momentum} to account for the presence of the
immersed boundary. The forcing term
\(\bm{f}\left( \bm{x},t \right)\) is the local body force
density at the fluid nodes (Eulerian nodes), it is distributed from
the surface force density \(\bm{F}\left( s,t \right)\)
(\(s\) is a parametrization of the immersed boundary surface) at the
immersed boundary points (Lagrangian points -- we use traditional terminology used in \citep{wu_shu_2009_implicit}). The forcing term is
expressed as:
\begin{equation}
    \bm{f}(\bm{x},t) = 
    \int_{\partial S} \bm{F}(\bm{s},t) \, \delta\left( \bm{x}-\bm{X}(s,t) \right)\mathrm{d}s,
\end{equation}
where \(\bm{x}\) and \(\bm{X}\left( s\bm{,}t \right)\)
denote the Eulerian nodes and Lagrangian point coordinates that discretize the
fluid domain and the immersed boundary, respectively. Eulerian nodes and
Lagrangian points interact through Dirac delta function
\(\delta\left( \bm{x} - \bm{X}\left( s,t \right) \right)\). Therefore,
the velocity at the immersed boundary points is calculated from the
velocity at the Eulerian nodes using the Dirac delta function
\begin{equation}
    \bm{U}(\bm{X}(s,t)) =
    \int_{\Omega} \bm{u}(\bm{x}) \, \delta\left( \bm{x}-\bm{X}(s,t) \right) \mathrm{d}V.
\end{equation}

\subsection{Incremental Pressure Correction Scheme (IPCS)}
\label{sec:ipcs}

The Incremental Pressure Correction Scheme (IPCS)
\citep{goda_1979_multistep}
is an operator splitting method,
which transforms the nonlinear Navier--Stokes (N-S)
equations into algebraic equations by coupling the pressure and
velocity field values (N-S equations are difficult to solve due to their
inherent non-linearity). IPCS is a modified version of the fractional
step method proposed by
\citet{chorin_1968_numerical}
and
\citet{temam_1969_approximation}. It improves the accuracy of the original scheme with little extra cost.
In our description we consider no external forces
for the N-S equations,
and in Section 2.3 we explain in detail how IB forces were incorporated.

The IPCS scheme involves three steps. In the first step, we compute a tentative velocity \(\bm{u}^{*}\) by
advancing the linear momentum Eq.\eqref{eq:momentum} in time using the backward
Euler difference scheme,
\(\dot{\bm{u}} = \frac{\bm{u}^{n+1} - \bm{u}^{n}}{\Delta t}\).
We linearize the nonlinear terms using a semi-implicit method, such that the advection term
\(\bm{u} \cdot \bm{\nabla}\bm{u}\) becomes
\(\bm{u}^{n} \cdot \bm{\nabla}\bm{u}^{n+1}\). Using this
semi-implicit approach for linearization the Courant-Friedrichs-Lewy (CFL) condition, which
limits the time step according to the velocity and
spatial discretization to ensure stability of the solution, becomes less restrictive.
Additionally, the kinematic viscosity is written as
\(v = v\left( \bm{u}^{n} \right)\), which simplifies the N-S
equations that are now written as a linearized set of equations, known
as the Oseen equations:
\begin{equation}
    \label{eq:oseen}
    \bm{u}^{n+1}
    + \Delta t\bm{u}^{n} \cdot \nabla\bm{u}^{n+1}
    - \Delta t\nabla \cdot 2v\bm{\varepsilon}(\bm{u}^{n+1})
    + \Delta t\nabla p^{n+1} 
    = \bm{u}^{n},
\end{equation}
\begin{equation}
    \nabla \cdot \bm{u}^{n+1} = 0.
\end{equation}

In Eq.\eqref{eq:oseen}, \(p^{n+1}\) is still unknown.
Therefore, we will compute the
tentative velocity \(\bm{u}^{*}\)
(\(\bm{u}^{*} \approx \bm{u}^{n+1}\)), by replacing in Eq.\eqref{eq:oseen}
\(p^{n+1}\) with the known value \(p^{n}\). 
This makes Eq.\eqref{eq:oseen} easier to solve for the tentative velocity
\(\bm{u}^{*}\):

\begin{equation}
    \label{eq:tentative-velocity}
    \bm{u}^{*} 
    + \Delta t\,\bm{u}^{n} \cdot \nabla\bm{u}^{n+1}
    - \Delta t\,\nabla \cdot 2v\bm{\varepsilon}\left( \bm{u}^{*} \right)
    + \Delta t\,\nabla p^{n} 
    = \bm{u}^{n}.
\end{equation}

Equation \eqref{eq:tentative-velocity} is numerically solved to compute the tentative velocity
\(\bm{u}^{*}\) without using the incompressibility constraint
Eq.\eqref{eq:incompressibility-constraint}.

In the second step, we use the tentative velocity
\(\bm{u}^{*}\) to compute the updated velocity \(\bm{u}^{n+1}\),
which is divergence-free (and should fulfill the incompressibility
constraint). We define a function for the velocity correction as
\(\bm{u}^{c} = \bm{u}^{n+1} - \bm{u}^{*}\), and after some
algebra (subtracting Eq.\eqref{eq:tentative-velocity} from Eq.\eqref{eq:oseen}) 
and using that
\(\bm{\nabla} \cdot \bm{u}^{n+1} = 0 \Rightarrow \ \bm{\nabla} \cdot \bm{u}^{c} = - \bm{\nabla} \cdot \bm{u}^{*}\)
(incompressibility constraint) we obtain a new set of flow equations for
\(\bm{u}^{c}\):
\begin{equation}
    \label{eq:uc1}
    \bm{u}^c 
    + \Delta t\,\bm{u}^{n} \cdot \nabla\bm{u}^c
    - \Delta t\,\nabla \cdot 2v\bm{\varepsilon}\left( \bm{u}^c \right)
    + \Delta t\,\nabla\Phi^{n+1}
    = 0,
\end{equation}
\begin{equation}
    \label{eq:uc2}
    \nabla \cdot \bm{u}^{c} = - \nabla \cdot \bm{u}^{*},
\end{equation}
where \(\Phi^{n+1} = p^{n+1} - p^{n}\). 
Equation \eqref{eq:uc1} can be further simplified to
\(\bm{u}^{c}\bm{+}\Delta\text{t\ }\bm{\nabla}\Phi^{n+1}\bm{= 0}\)
and the system of Eqs.(10)-(11) is reduced to an elliptic type (Poisson)
partial differential equation for the pressure difference \(\Phi^{n}\)
\begin{equation}
    \nabla^2 \Phi^{n+1} = \frac{\nabla \cdot \bm{u}^*}{\Delta t},
\end{equation}
with boundary conditions being those applied to the original flow
problem.

In the third and final step, we compute the updated velocity
\(\bm{u}^{n+1}\) and pressure \(p^{n+1}\) through
\(p^{n+1} = \Phi^{n+1} + p^{n}\) and
\(\bm{u}^{n+1} = \bm{u}^{*} + \Delta t\bm{\nabla}\Phi^{n+1}\),
respectively. 

The algorithmic procedure for our implementation of the IPCS for solving N-S equations is summarized below:

\begin{enumerate}
\item
  Calculate the tentative velocity \(\bm{u}^{*}\) by solving Eq.\eqref{eq:tentative-velocity}
\item
  Solve the Poisson equation
  \(\bm{}\nabla^2 \Phi^{n+1}\bm{=}\frac{\nabla \cdot \bm{u}^{*}}{\Delta t}\)
  (Eq.(12))
\item
  Calculate the corrected velocity
  \(\bm{u}^{n+1} = \bm{u}^{*} + \Delta t\bm{\nabla}\Phi^{n+1}\)
  and pressure \(p^{n+1} = \Phi^{n+1} + p^{n}.\)
\end{enumerate}

\subsection{Boundary Condition Enforced Immersed Boundary (BCE-IB) method}
\label{sec:bce-ibm}

In this section, we describe the boundary condition enforced immersed
boundary (BCE-IB) method for the numerical solution of the
incompressible 3D Navier--Stokes equations. 

The immersed boundary method has two steps:

\begin{itemize}
\item
  \emph{Predictor step}: where we numerically solve the N-S equations to
  compute the predicted velocity field
  \(\tilde{\bm{u}}\left( \bm{x} \right)\) by disregarding
  the body force terms in Eq.(1) (\(\bm{f} = \bm{0}\))
\end{itemize}

\begin{equation}
   \rho\frac{\tilde{\bm{u}}-\bm{u}^{n}}{\Delta t}
   + \rho\left( \bm{u}^n \cdot \nabla\bm{u}^n \right) 
   = + \Delta t\nabla p^{n} 
   - \Delta t\nabla \cdot 2v\bm{\varepsilon}(\bm{u}^{n})
\end{equation}

\begin{itemize}
\item
  \emph{Corrector step}: where we account for the effect of body forces
  and update the predicted velocity field \(\tilde{\bm{u}}\) to
  the updated one \(\bm{u}^{n + 1}\)\textbf{,} which satisfies the
  boundary condition
  \(\bm{u}\left( \bm{X}\left( s \right),t \right) = \bm{U}_{B}\)
\end{itemize}

\begin{equation}
    \rho \frac{\bm{u}^{n + 1} - \tilde{\bm{u}}}{\Delta t} 
    = \bm{f}^{n + 1}.
\end{equation}

In the predictor step, Eq.(13) is used to compute the predicted velocity
field \(\tilde{\bm{u}}\) under the incompressibility constraint
(mass conservation), which couples the velocity and pressure. The
predictor step accounts for the first of the three steps
in the IPCS method. In this step, we compute the predicted velocity
\(\tilde{\bm{u}}\), which in this case is the tentative velocity
\(\bm{u}^{*}\) (Step 1 in the IPCS algorithmic procedure). The
corrector step, involves the evaluation of the unknown body force
\(\bm{f}^{n + 1}\) and the update of the predicted velocity
\(\tilde{\bm{u}}\) to the physical one
\(\bm{u}^{n + 1}\) (Steps 2 and 3 in the IPCS algorithmic
procedure).

Accurate and efficient computation of the unknown body force \(\bm{f}^{n + 1}\) is an important
advantage of the Boundary Condition Enforced Immersed Boundary (BCE-IB) method. In this method,
the body force \(\bm{f}^{n + 1}\) is equivalent to a velocity
correction, which applies implicitly such that the velocity
\(\bm{U}\left(\bm{X}\left( s \right),t \right)\) on the
Lagrangian points, interpolated from the physical velocity
\(\bm{u}\left( \bm{x},t \right)\) computed on the Eulerian nodes, equals
to the prescribed boundary velocity \(\bm{U}_{B}\) (Eq.~\eqref{eq:noslip_bc}). The
body force term \(\bm{f}^{n + 1}\) is computed using the equation
\begin{equation}
    \bm{U}_B^{n+1}\left( \bm{X}^{n+1} \right) =
    \int_{\Omega} \left( \tilde{\bm{u}} + \Delta t\frac{\bm{f}^{n + 1}}{\rho} \right)
    \delta\left( \bm{x} - \bm{X}^{n+1} \right) \mathrm{d}V,
\end{equation}
derived by substituting Eq.(14) to Eq.(6). The force density
\(\bm{f}^{n + 1}\left( \bm{x},t \right)\), is evaluated on the
Eulerian nodes, and it is computed by distributing (`spreading') the
boundary force \(\bm{F}\left( \bm{X}\left( s \right),t \right)\)
on the Lagrangian points through the Dirac delta function
\(\delta\left( \bm{x - X}\left( s,t \right) \right)\) (see
Eq.(5)). Eq.(15) is written as
\begin{equation}
    \bm{U}_B^{n+1} \left( \bm{X}^{n+1} \right) = 
    \int \left( \tilde{\bm{u}} + \Delta t\frac{\int {\bm{F}^{n+1} 
    \left( \bm{X}^{n + 1} \right)
    \delta\left( \bm{x} - \bm{X}^{n+1} \right)\mathrm{d}s}}{\rho} \right)
    \delta\left( \bm{x} - \bm{X}^{n+1} \right)\mathrm{d}V,
\end{equation}
and \(\bm{U}_{B}^{n + 1}\) is no longer correlated with
\(\bm{f}^{n + 1}\) but instead with \(\bm{F}^{n + 1}\).

To compute the boundary force \(\bm{F}^{n + 1}\), we rewrite Eq.(16)
in a discrete form that results in an algebraic system of equations.
We represent the immersed boundary by a set of Lagrangian
points
\(\bm{X}_{i} = \left( X_{i},Y_{i},Z_{i} \right),\ i = 1,2,\ldots,M\)
and the flow domain by the Eulerian nodes
\(\bm{x}_{j} = \left( x_{j},y_{j},z_{j} \right),\ j = 1,2,\ldots,N\).
In our analysis,
the Eulerian mesh in the vicinity of the immersed object
has a uniform Cartesian grid structure
with grid spacing \(h\).
Furthermore, the Dirac function
\(\delta\left( \bm{x} - \bm{X}\left( s,t \right) \right)\) is
approximated by a continuous kernel distribution
\(D\left( \bm{x}_{i} - \bm{X}^{j} \right)\) given as
\begin{equation}
    D_{ij} = D\left( \bm{x}_i - \bm{X}^j \right) =
    \left( \frac{1}{h}\delta\left( \frac{x_i - X^j}{h} \right) \right)
    \left( \frac{1}{h}\delta\left( \frac{y_i - Y^j}{h} \right) \right)
    \left( \frac{1}{h}\delta\left( \frac{z_i - Z^j}{h} \right) \right),
\end{equation}
with the kernel \(\delta\left( r \right)\) proposed by
\citet{lai_peskin_2000_immersed}
\begin{equation}
    \delta(r) = 
        \begin{cases}
        \begin{alignedat}{2}
            \frac{1}{8}\left( 3 - 2|r| + \sqrt{1 + 4 |r| - 4r^2} \right),   &\quad&     |r| & \leq 1 \\
            \frac{1}{8}\left( 5 - 2 |r| - \sqrt{-7 + 12|r| - 4r^2} \right), &\quad& 1 < |r| & \leq 2 \\
            0,                                                              &\quad&     |r| & > 2
        \end{alignedat}
        \end{cases}
\end{equation}

Using Eq.(6) for the force term 
\begin{equation}
    \bm{f}^{n+1}(\bm{x}_j) =
    \sum_{i=1}^M {\bm{F}^{n+1}\left( \bm{X}_i^{n+1} \right)
    D_h^{ij}\Delta S_i},
\end{equation}
Eq.(16) is written as
\begin{equation}
   \bm{U}_{B}^{n+1}(\bm{X}_i^{n+1}) 
   = \sum_{j=1}^N \tilde{\bm{u}}(\bm{x}_j) D_h^{ij} h^3 
   + \sum_{j=1}^N \sum_{k=1}^M \frac{\bm{F}^{n+1}(\bm{X}_i^{n+1}) \Delta t}{\rho}
   D_h^{kj} D_h^{ij} h^3,
\end{equation}
where \(\Delta S_{i}\) is the area of the \emph{i\textsuperscript{th}}
surface element (segment). Eq.(20) forms a well-defined system of
equations for the variables
\(\bm{F}_i^{n + 1}\bm{\ }\left( i = 1,2,\ldots,M \right)\).
Eq.(20) can be written in matrix notation as
\begin{equation}
    \bm{A}_{\bm{F}}\bm{F} = \bm{B}_{\bm{F}},
\end{equation}
with \(\bm{A}_{\bm{F}},\bm{F}\) and
\(\bm{B}_{\bm{F}}\) defined as
\begin{equation}
    \bm{A}_{\bm{F}} =
    \frac{\Delta t}{\rho}h^3
    \begin{bmatrix}
        D_{11} \Delta S_1 & D_{12} \Delta S_1 & \dots  & D_{1N} \Delta S_1 \\
        D_{21} \Delta S_2 & D_{22} \Delta S_2 & \dots  & D_{2N} \Delta S_2 \\
        \vdots            & \vdots            & \ddots & \vdots            \\
        D_{M1} \Delta S_M & D_{M2} \Delta S_M & \dots  & D_{MN} \Delta S_M
    \end{bmatrix}
    \begin{bmatrix}
        D_{11} & D_{12} & \dots  & D_{1M} \\
        D_{21} & D_{22} & \dots  & D_{2M} \\
        \vdots & \vdots & \ddots & \vdots \\
        D_{N1} & D_{N2} & \dots  & D_{NM}
    \end{bmatrix},
\end{equation}
\begin{equation}
    \bm{B}_{\bm{F}} =
    \bm{U} - h^3 \bm{D}^\mathrm{T} \tilde{\bm{u}} =
    \begin{bmatrix}
        \bm{U}_1 \\
        \bm{U}_2 \\
        \vdots       \\
        \bm{U}_M 
    \end{bmatrix}
    - h^3
    \begin{bmatrix}
        D_{11} & D_{21} & \dots  & D_{N1} \\
        D_{12} & D_{22} & \dots  & D_{N2} \\
        \vdots & \vdots & \ddots & \vdots \\
        D_{1M} & D_{2M} & \dots  & D_{NM}
    \end{bmatrix}
    \begin{bmatrix}
        \tilde{\bm{u}}_1 \\
        \tilde{\bm{u}}_2 \\
        \vdots               \\
        \tilde{\bm{u}}_N
    \end{bmatrix},
\end{equation}

\begin{equation}
    \bm{F} =
    \begin{bmatrix}
        \bm{F}_1 \\
        \bm{F}_2 \\
        \vdots       \\
        \bm{F}_M 
    \end{bmatrix},
\end{equation}
where \(\bm{U}_i\ (i\bm{=}1,2,\ldots,M\bm{)}\),
\(\bm{F}_i\bm{\ (}i\bm{=}1,2,\ldots,M\bm{)}\)
and
\({\tilde{\bm{u}}}_j\bm{\ }\left( j = 1,2,\ldots,M \right)\)
are the abbreviations for
\(\bm{U}_{B}^{n + 1}\left( \bm{X}_{\bm{i}}^{n + 1} \right)\),
\(\bm{F}^{n + 1}\left( \bm{X}_{\bm{i}}^{n + 1} \right)\) and
\(\tilde{\bm{u}}\left( \bm{x}_{j} \right)\), respectively.
By solving he system of equations (Eq.(22)) using a direct solver, we obtain the unknown boundary force
\(\bm{F}_{\bm{i}}^{n + 1}\bm{\ }\left( i = 1,2,\ldots,M \right)\)
at all Lagrangian points. Boundary forces
\(\bm{F}_{\bm{i}}^{n + 1}\) are then substituted into Eq.(19)
and Eq.(16) to calculate the body force \(\bm{f}^{n + 1}\) on the Eulerian nodes, and the
corrected physical velocity \(\bm{u}^{n + 1}\)\emph{. }

\subsection{Algorithmic procedure}

Our approach combines the IPCS (section~\ref{sec:ipcs})
with the BCE-IB method (section~\ref{sec:bce-ibm}).
The velocity field predicted by IPCS 
is corrected to account for the force contribution
from the immersed boundary.
The proposed solution procedure can be summarized as follows;
to update the solution from time instance \(n\) to \(n+1\):


\begin{enumerate}

\item
  Calculate the tentative velocity \(\tilde{\bm{u}}\) using Eq.(13)
\item
  Compute the matrix \(\bm{A}_{\bm{F}}\) using Eq.(22)
\item
  Solve the system (Eq.(21)) to compute the boundary force
  \(\bm{F}^{n + 1}\left( \bm{X}_i^{n + 1} \right)\bm{,\ (}i\bm{=}1,2,\ldots,M\bm{)}\)
  on the Lagrangian points. Substitute the computed boundary force
  \(\bm{F}^{n + 1}\left( \bm{X}_i^{n + 1} \right)\)
  into Eq.(19) to obtain the body force
  \(\bm{f}^{n + 1}\left( \bm{x}_{j}^{n + 1} \right)\bm{,\ (}j\bm{=}1,2,\ldots,N\bm{)}\) on the Eulerian points.
\item
  Use \(\tilde{\bm{u}}\) to update the corrected velocity
  \(\bm{u}^{(n+1)}=\frac{\Delta t \bm{f}^{n+1} + \tilde{\bm{u}}}{\rho}\)
  using Eq.(14)
 \item
  Solve the Poisson equation
  \(\bm{}\nabla^2 \Phi^{n+1}\bm{=}\frac{\nabla \cdot \bm{u}^{n+1}}{\Delta t}\) for $\Phi^{n}$
  (Eq.(12))
\item
  Recompute the updated velocity
  \(\bm{u}^{n+1} = \bm{u}^{n} + \Delta t\bm{\nabla}\Phi^{n+1}\)
  and pressure \(p^{n+1} = \Phi^{n+1} + p^{n}.\)
\item
  Set physical/updated velocity \(\bm{u}^{\left( n + 1 \right)}\) as \(\bm{u}^{\left( n  \right)}\) and repeat steps (1) to (4) until the desired
  solution is achieved.
\end{enumerate}

Steps 1, 5, 6 apply to the IPCS method without the presence of the immersed boundary, while Steps 2, 3, 4 calculate the velocity field (on the Eulerian nodes) due to the presence of the immersed boundary.
\subsection{Surface area for Lagrangian points}

In 3D flow cases, the surface area \({\Delta S}_{i}\) that is assigned
to each Lagrangian point is not as straightforward to define as in two
dimensions. Immersed boundaries are represented by a set of Lagrangian
points. Therefore, the number of these points and the way they are
distributed over the immersed boundary directly affect the accuracy of
the IB method.

In 2D, the distribution and number of Lagrangian points are always
investigated through the ratio of \(\frac{\text{ds}}{h}\), with
\(\text{ds}\) being spacing of the Lagrangian points and \(h\) the
Eulerian nodal spacing. High ratio of \(\frac{\text{ds}}{h}\) will cause
fluid leakage (spreading from Lagrangian points to Eulerian nodes and
vice-versa is not accurate), while low ratio will increase the
computational cost (as rows in matrix
\(\bm{A}_{\bm{F}}\) in Eq.(21) will be linearly dependent and
the matrix will become ill-conditioned and therefore difficult to
invert). In the literature, there are many
studies which report on selection of \(\frac{\text{ds}}{h}\);
\citet{uhlmann_2005_immersed}
suggests that Lagrangian points spacing should be equal
to the Eulerian mesh/grid resolution \(\frac{\text{ds}}{h}=1\),
\citet{su_etal_2007_immersed}
and
\citet{kang_hassan_2011_comparative}
used \(\frac{\text{ds}}{h}=1.5\) (Lagrangian points spacing is larger than
the Eulerian mesh/grid resolution), while
\citet{limaesilva_etal_2003_numerical}
reported \(\frac{\text{ds}}{h}\leq 0.9\) (Lagrangian points spacing
is less than the Eulerian nodal resolution). In 3D, the surface area
\({\Delta S}_{i}\) appears as a key feature of the IB method. In
this study, we propose and use an easy to apply method to compute
\({\Delta S}_{i}\). The area of each surface element \({\Delta s}_{i}\)
should be close to \((ah)^{2}\), with \(0.1 \leq a \leq 2.5\) (just like
having in 2D \(0.3 \leq \frac{ds}{h} \leq 2.5\), with \(h\) being the
spacing of the Eulerian nodes).

In this study, we uniformly distribute the Lagrangian points to
assign the same value \(\Delta S\) to all Lagrangian points. To obtain a
distribution of Lagrangian points with uniform surface element area, we
generate surface mesh (the nodes are the Lagrangian points used in the IB
method) with element edges having lengths with small standard deviation. The area element \(\Delta S\) for each Lagrangian point is the
computed mean value of the facets area.

\subsection{Derived quantities---wall shear stress (WSS)}

An important aspect of blood flow simulations is the numerical
computation of flow parameters that are of clinical importance, such as
blood vessel wall shear stress (WSS). While focusing on WSS, we present an accurate meshless point collocation to compute derived quantities from the velocity field obtained from blood flow simulations. Meshless methods (MMs) have been well-established for the efficient and accurate representation of 3D complex geometries as point clouds, and for the accurate computation of spatial derivatives. In our study, we use the Discretization Corrected Particle Strength Exchange (DC PSE) \citep{DCPSE}
meshless method. The DC PSE method, is one of the most
accurate meshless point collocation methods to compute spatial
derivatives of field functions defined on point clouds.

WSS is defined as the magnitude of the surface traction vector \(\bm{t}_{\bm{s}}=\bm{t}-(\bm{t} \cdot \hat{\bm{n}}) \cdot \hat{\bm{n}}\), where \(\hat{\bm{n}}\) is the unit normal vector pointing outward, and \(\bm{t} =2\mu{\bm{\varepsilon}}\) is the shear stress, defined
based on the fluid's viscosity \(\mu\) and the rate of deformation
tensor \({\bm{\varepsilon}}\). Let
\(\bm{u}\left( \bm{x},t \right)\) be continuously differentiable
velocity vector field describing the blood flow at a point in the domain
of interest. The velocity field values are computed by interpolating the
velocity field computed on the Cartesian flow domain using the IB
method, on a body-fitted mesh of the immersed object (the mesh is used only for visualization of the flow field and derived quantities such as WSS).We consider a
discretization of the domain of interest (immersed object) using a finite-element mesh
(tetrahedral linear elements), and denote the corresponding set of nodes
by \(N\). The nodal subset \(N_{S}\) with points on the surface of the
tetrahedral mesh will be used for the calculation of WSS. Using indicial notation for the
components of \(\bm{u}\), the deformation (strain-rate) tensor
components are defined by
\({{\varepsilon}}_{{ij}} = \frac{1}{2}\left( u_{i,j} + u_{j,i} \right)\) for
\(i,j = \left\{ x,y,z \right\}\), where
\(u_{i,j} = \frac{\partial u_{i}}{\partial x_{j}}\). The
terms \(\frac{\partial u_{i}}{\partial x_{j}}\) of
the deformation tensor \({\bm{\bm{\varepsilon}}}\) are computed on the surface points \(N_{S}\) using the DC PSE method as
\begin{equation}
    \frac{\partial u_{i}}{\partial x_{j}} = \Phi_{,j}u_{i},
\end{equation}
with \(\Phi_{,j}\) being the \emph{j}\textsuperscript{th} spatial derivative
computed, and \(u_{i}\) being a column vector of
the \emph{i}\textsuperscript{th} velocity vector components. 

\section{Algorithm verification}

We demonstrate the accuracy, efficiency and applicability of our method
on internal flow cases. The numerical results obtained were compared
against analytical solutions (where available) and experimental data. We consider two 3D benchmark flown problems, the Poiseuille flow in a
straight tube and the angle tube bend flow problem, where
an analytical solution (Poiseuille flow) and experimental data (tube
bend) are available, respectively.

\subsection{Poiseuille flow in a cylindrical tube}

We demonstrate the accuracy of the proposed IB method considering the
Poiseuille flow in a straight cylindrical tube. The tube has length \(L = 10R\) and radius \(R = 0.005\)
\emph{m}, and is immersed in a domain (which is the
spatial domain where the flow equations are solved) with dimensions
\(- 0.006 \ m \leq y,z \leq 0.006 \ m\) and \(0 \ m \leq x \leq 0.05 \ m\), as
shown in Fig.~1. The immersed cylinder is described by a set of points on
its surface (Lagrangian points). The box
domain is discretized with a high quality tetrahedral mesh. The vertices of the tetrahedral elements (Eulerian nodes) form a uniform Cartesian grid. To generate the mesh we use the FEniCS build-in function BoxMesh \cite{unknown_2019_fenics}.

\begin{figure}
    \centering
    \includegraphics[width=\textwidth]{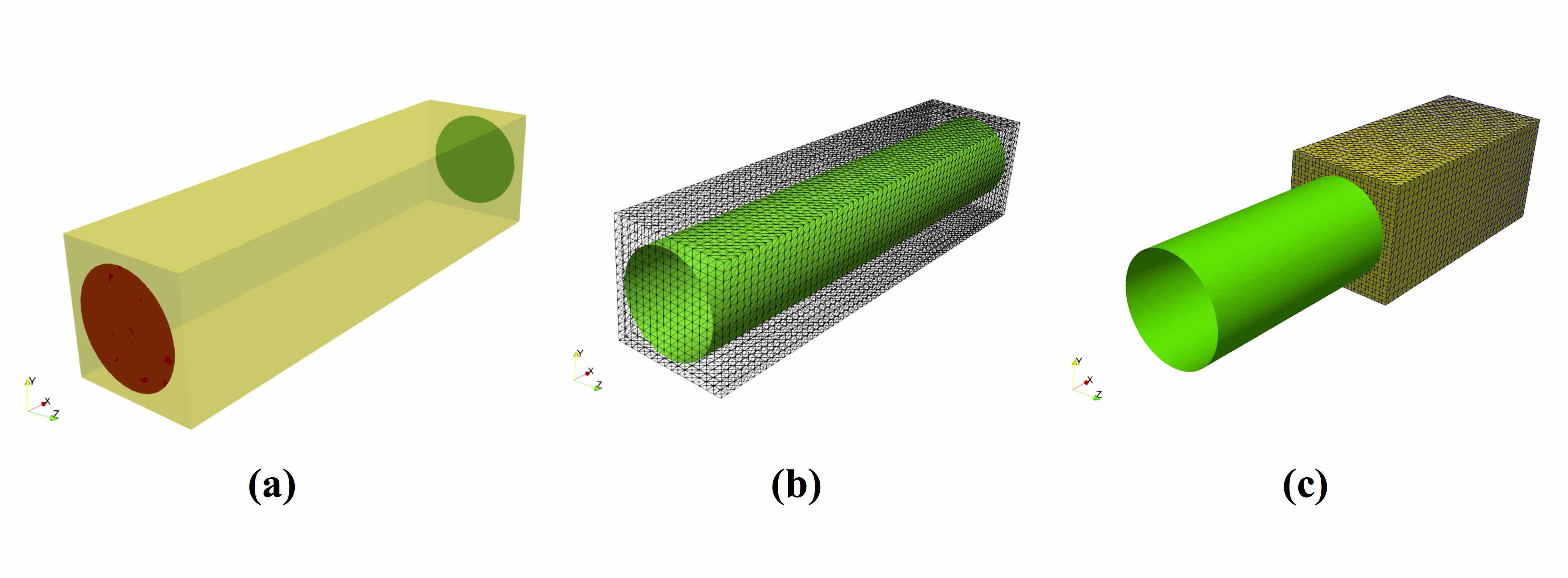}
    \caption{
        \textbf{(a)}
        Flow domain for the Poiseuille flow example with 
        inlet (circle in red color),
        outlet (circle in green color) and
        walls (yellow color) boundaries;
        \textbf{(b)}
        wireframe of the tetrahedral mesh
        and the immersed boundary (Lagrangian points);
        \textbf{(c)}
        vertical cross-section of the tetrahedral mesh
        along with the immersed boundary.
    }
    \label{fig:my_label}
\end{figure}

The driving force of the flow is the pressure difference applied to the
inlet and outlet. 
In our simulations, the density of the fluid was set to
\(\rho = 1,050~\mathrm{kg\,m^{-3}}\), and the dynamic viscosity to
\(\mu = 0.00345~\mathrm{N\,s/m^{2}}\).
At the inlet (red circle in Fig.~1a) and
outlet (green circle in Fig.~1a), we apply pressure boundary
conditions, while at the remaining walls (surface yellow Fig.~1a), we
apply no-slip velocity boundary conditions (\(\bm{u} = 0\)). The time step was set to
\(dt = 2.5 \times 10^{- 3}\). The simulation ends when the normalized
root mean square difference
\begin{equation}
    L_{\text{NRMSE}} = \frac{\sqrt{\frac{\sum_{i = 1}^{N}\left( u_{i}^{t + dt} - u_{i}^{t} \right)^{2}}{N^2}}}{\left( \max\left( u_{i}^{t} \right) - \min\left( u_{i}^{t} \right) \right)},
\end{equation}
between two successive time instances for all three velocity components
is less than \(10^{-8}\)
(practically, the flow reaches a steady state). The exact solution for the velocity field components is
\(u_{x} = u_{y} = 0\) and
\(u_{z}\left( r \right) = U_{\max}\left( 1 - \left( \frac{r}{R} \right)^{2} \right)\),
with \(U_{\max} = - \frac{R^{2}}{4\mu}\frac{\text{dp}}{\text{dz}}\)
being the maximum velocity. By setting the pressure difference to
\(\frac{\text{dp}}{\text{dz}} = \frac{1}{L}\) (applying
\(p_{\text{inlet}} = 1\) and \(p_{\text{outlet}} = 0\)) the Reynolds
number becomes \(\mathrm{Re} = \frac{U_{\max}L}{\mu} \approx 551\), while for
\(p_{\text{inlet}} = 2\) Reynolds number becomes
\(\mathrm{Re} = \frac{U_{\max}L}{\mu} \approx 1,102\).
\begin{table}
    \centering
    \caption{
        Tetrahedral mesh resolution, 
        number of Eulerian nodes,
        number of tetrahedral elements, and
        number of Lagrangian points
        for the successively denser grids 
        considered for the Poiseuille flow problem.
    }
    \begin{tabular*}{\textwidth}
    {@{\extracolsep{\fill}} cccc }
        \toprule
        Grid resolution \(h\)~(m) & Eulerian nodes & Tetrahedral elements & Lagrangian pointss \\
        \midrule
        $1   \times 10^{-3}$  &    8,619  &     43,200  &   1,632 \\
        $5   \times 10^{-4}$  &   63,125  &    345,600  &   6,464 \\
        $2.5 \times 10^{-4}$  &  482,601  &  2,764,800  &  25,527 \\
        \bottomrule
    \end{tabular*}
    \label{tab:poiseuille-grids}
\end{table}

We use successively denser tetrahedral meshes to obtain a mesh independent numerical solution. 
The meshes are described in Table~\ref{tab:poiseuille-grids}.
Table~\ref{tab:poiseuille-norms},
reports the maximum absolute error
\(L_{\infty} = \underset{i}{\max} \left| u_{i}^{\text{numerical}} - u_{i}^{\text{exact}} \right|\)
and the root mean square error (RMSE)
\(L_{2} = \frac{1}{N}\sqrt{\sum_{i = 1}^{N}\left( u_{i}^{\text{numerical}} - u_{i}^{\text{exact}} \right)^{2}}\) for different mesh resolution, for \(\mathrm{Re} = 551\) and
\(\mathrm{Re} = 1,102\). For \(\mathrm{Re} = 551\) the convergence rates for the
\(L_{\infty}\) and \(L_{2}\) error norms are 1.29 and 2.79,
respectively. The convergence rates for \(L_{\infty}\) and \(L_{2}\)
error norms do not change, demonstrating that the relative error does not change.

\begin{table}
    \centering
    \caption{
        Maximum absolute \(L_{\infty}\) and
        root mean square \(L_{2}\) error norms 
        for the numerical solution computed using the proposed IB method 
        and the analytical solution for the Poiseuille flow example using
        \(\mathrm{Re} = 551\) and \(\mathrm{Re} = 1,102\).
    }
    \begin{tabularx}{\textwidth}{@{}YYYY@{}}
        \toprule
        \multicolumn{2}{c}{$\mathrm{Re} = 551$} & \multicolumn{2}{c}{$\mathrm{Re} = 1,102$} \\
        \cmidrule(lr){1-2} \cmidrule(lr){3-4}
        $L_\infty$         & $L_2$              & $L_\infty$         & $L_2$ \\
        \midrule
        $1.47\times10^{-2}$ & $1.99\times10^{-4}$ & $2.95\times10^{-2}$ & $3.99\times10^{-4}$ \\
        $6.62\times10^{-3}$ & $3.07\times10^{-5}$ & $6.62\times10^{-3}$ & $6.18\times10^{-5}$ \\
        $2.44\times10^{-3}$ & $4.15\times10^{-6}$ & $4.89\times10^{-3}$ & $8.42\times10^{-6}$ \\
        \bottomrule
    \end{tabularx}
    \label{tab:poiseuille-norms}
\end{table}

Figure 2 shows the \(L_{\infty}\) and \(L_{2}\) error norms versus the
mesh resolution of the Eulerian mesh. For both \(\mathrm{Re} = 55\)1 and
\(\mathrm{Re} = 1,102\), both \(L_{\infty}\) and \(L_{2}\) error norms, the results suggest that the proposed scheme is accurate.

\begin{figure}
    \centering
    \includegraphics[width=0.8\textwidth]{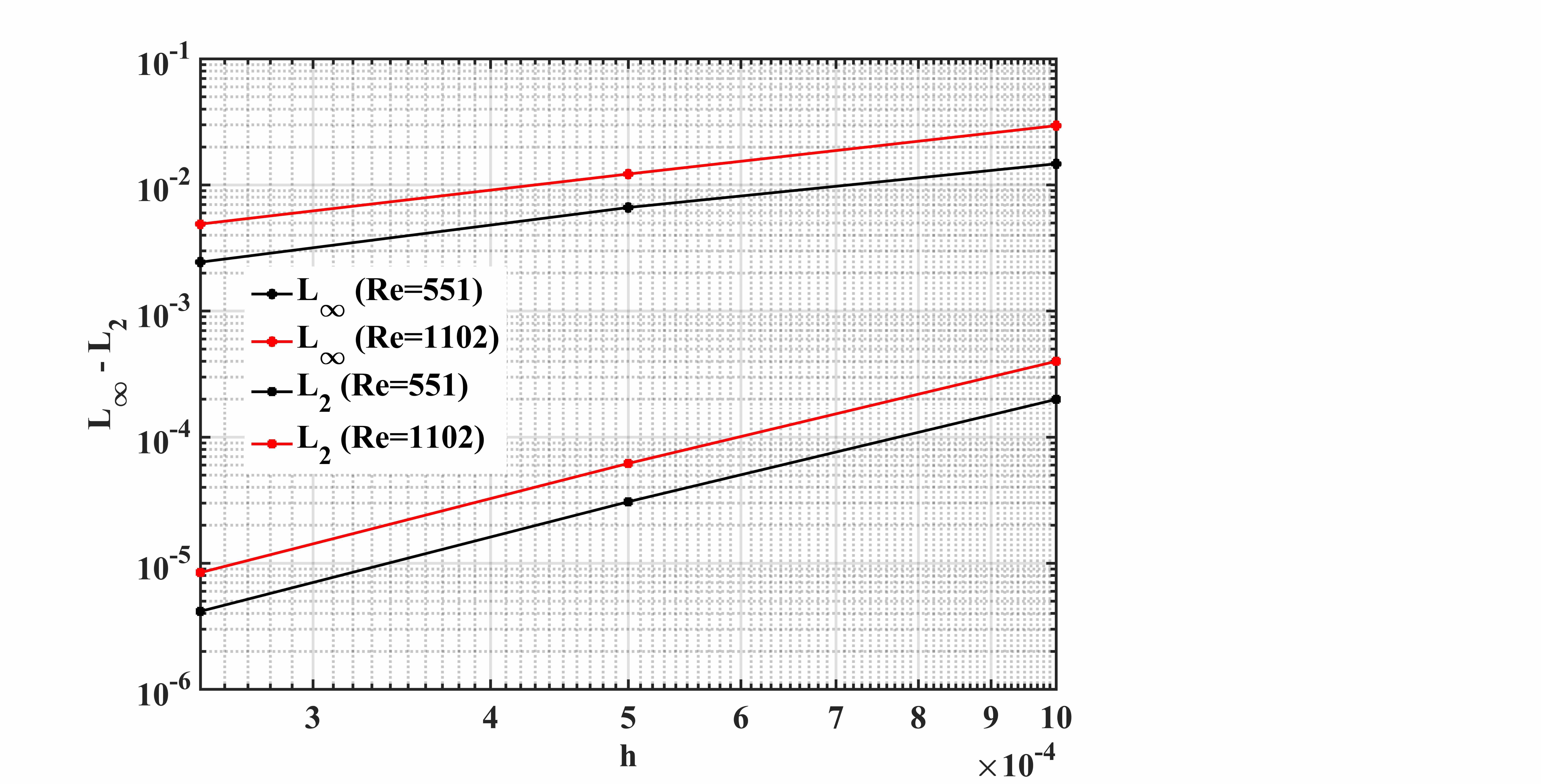}
    \caption{
        Maximum absolute \(L_{\infty}\) (red line) and
        root mean square \(L_{2}\) (black line) 
        error norms versus Eulerian tetrahedral mesh resolution \(h\) for the Poiseuille flow example.
    }
    \label{fig:my_label}
\end{figure}

Figure 3 shows the velocity streamlines for \(\mathrm{Re} = 551\) and \(\mathrm{Re} = 1,102\). The streamlines were produced on a tetrahedral body-fitted mesh (we refer to as visualization mesh) that consist of 345,600 linear elements. The velocity field computed with the IB method on the Eulerian nodes has been projected on the visualization mesh for visualization purposes (the visualization mesh is not used in the computation of the velocity and pressure field).

\begin{figure}
    \centering
    \includegraphics[width=\textwidth]{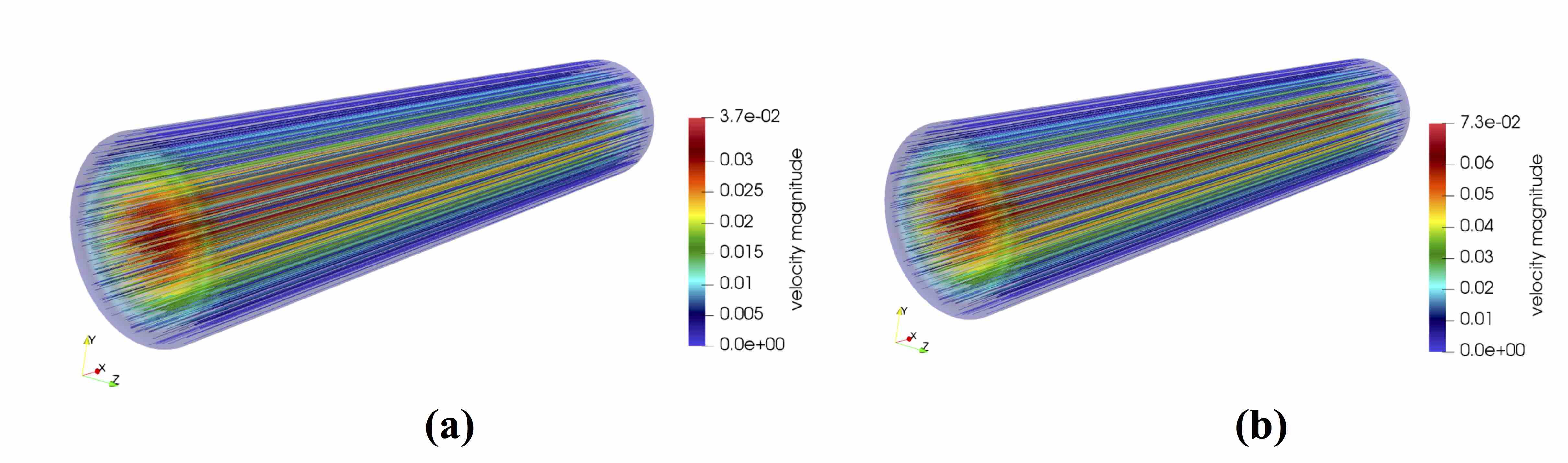}
    \caption{
        Velocity streamlines for \textbf{(a)} \(\mathrm{Re} = 551\) and \textbf{(b)} \(\mathrm{Re} = 1,102\) 
        using a visualization mesh of 345,600 linear elements for the Poiseuille flow example. The maximum velocity computed for \(\mathrm{Re} = 551\) and \(\mathrm{Re} = 1,102\) was 0.0361 and 0.0723, respectively (the theoretical values are 0.0362 and 0.0724, respectively)
    }
    \label{fig:my_label}
\end{figure}

\subsection{3D angle tube bend}

We further demonstrate the accuracy of the proposed method, considering
the flow in a three-dimensional 90° angle tube bend, as shown in Fig.~4.
We use this flow example to demonstrate the ability of the proposed
method to accurately capture flows in tortuous tubes where secondary
flows can occur. For U-bend, flow data are available;
\citet{vandevosse_etal_1989_finite}
performed laser Doppler velocimetry experiments to
obtain the center-plane axial velocities at a set of different angles
around the tube bend. To reduce disturbances in the flow field due to
inflow boundary conditions, we extended the inlet and outlet by 1 mm. In our numerical simulations, we
set the Reynolds number---computed based on the tube inner diameter
\(D_{i}\) and the mean inlet velocity \(\overline{U}\)---to
\(\mathrm{Re} = 300\), to match the experimental conditions. The inner diameter
\(D_{i}\) and the curvature radius \(R\) of the tube were set to 4 and
24 mm, respectively.

\begin{figure}
    \centering
    \includegraphics[width=0.8\textwidth]{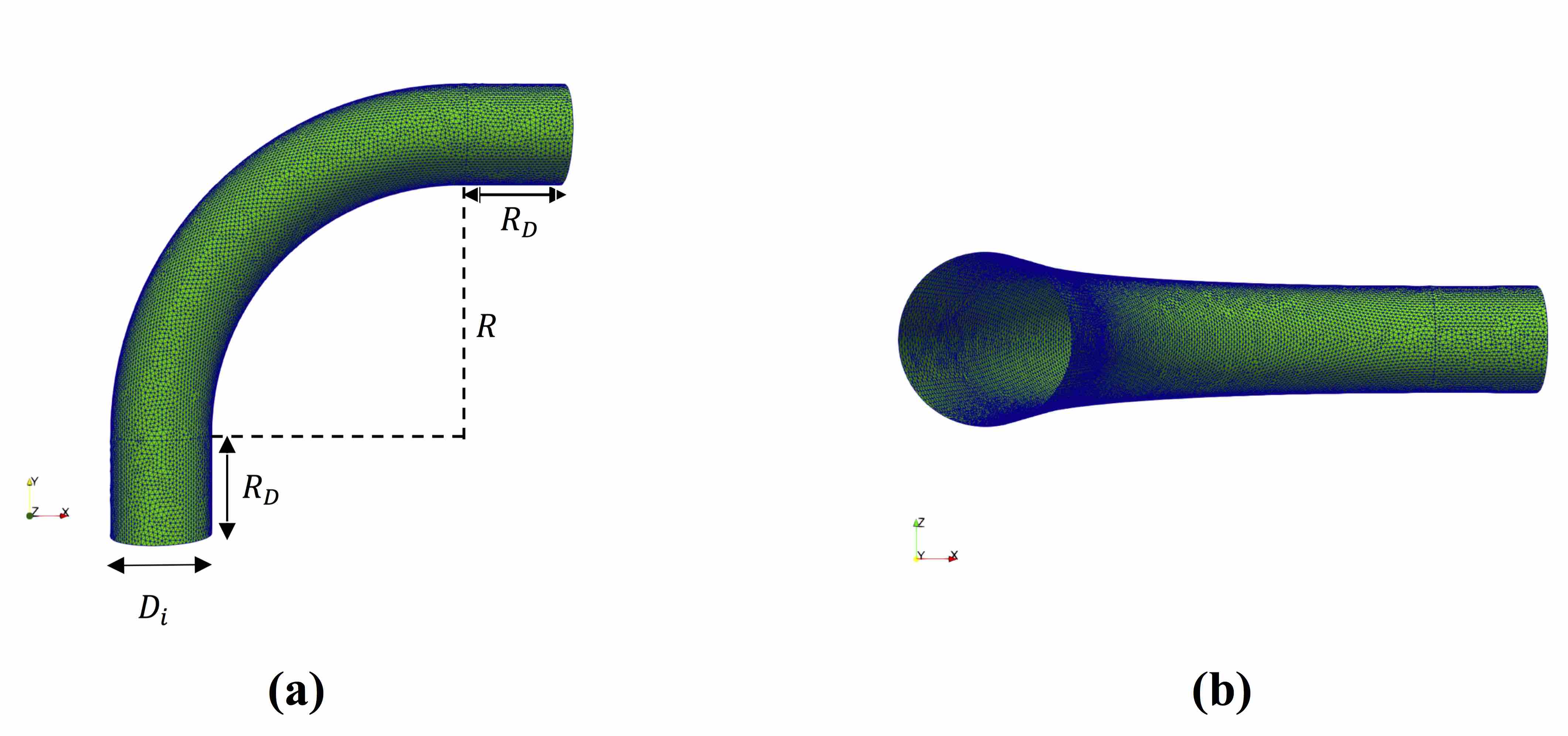}
    \caption{
        U-bend geometry
        \textbf{(a)} top and
        \textbf{(b)} side view.
        The points on the surface define the Lagrangian points of the immersed boundary $R_{D}=1 \ mm$, $R=24 \ mm$, $D_{i}=4 \ mm$.
    }
    \label{fig:my_label}
\end{figure}

The U-bend geometry is embedded within a box domain with
dimensions
\(- 0.0064 \ m \leq x \leq 0.032 \ m\), \(0 \ m \leq y \leq 0.0384 \ m\) 
and
\(- 0.0064 \ m \leq z \leq 0.0064 \ m\), as shown in
Fig.~5b. We discretize the flow domain using tetrahedral elements. The box
domain is discretized with a high quality tetrahedral mesh (the vertices of the tetrahedral elements are the Eulerian nodes), generated using a uniform Cartesian grid (to generate the mesh we used the FEniCS build-in function BoxMesh).

\begin{figure}[t]
    \centering
    \includegraphics[width=\textwidth]{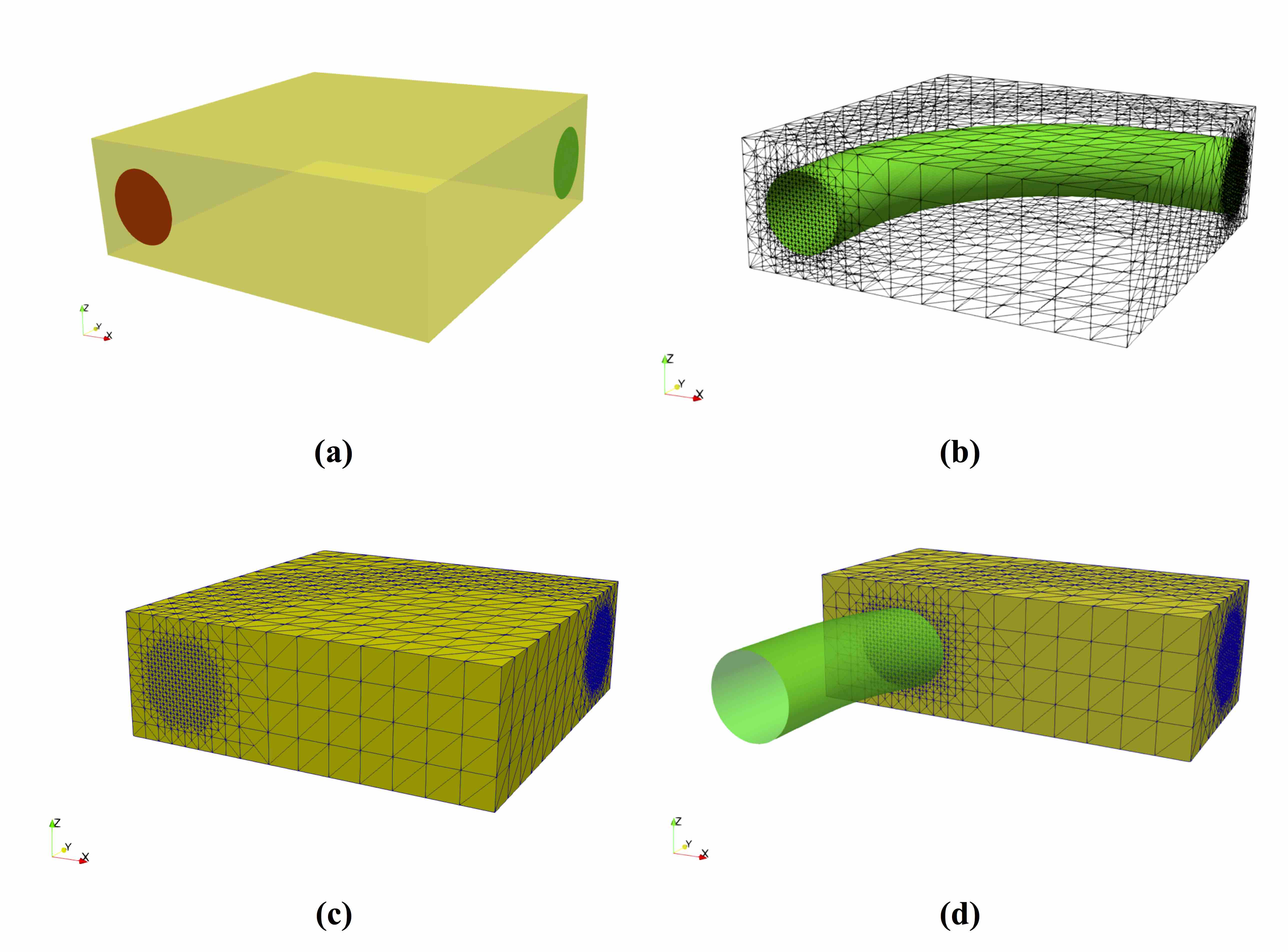}
    \caption{
        \textbf{(a)}
        Flow domain for the U-bend flow example, with 
        inlet (circle in red color), 
        outlet (circle in green color) and
        walls (yellow color) boundaries; 
        \textbf{(b)}
        wireframe of the tetrahedral mesh 
        and the U-bend immersed boundary (Lagrangian points);
        \textbf{(c)}
        tetrahedral mesh refined close and inside the immersed boundary, and
        \textbf{(d)}
        vertical cross-section of the tetrahedral mesh.
    }
    \label{fig:my_label}
\end{figure}

\begin{figure}
    \centering
    \includegraphics[width=\textwidth]{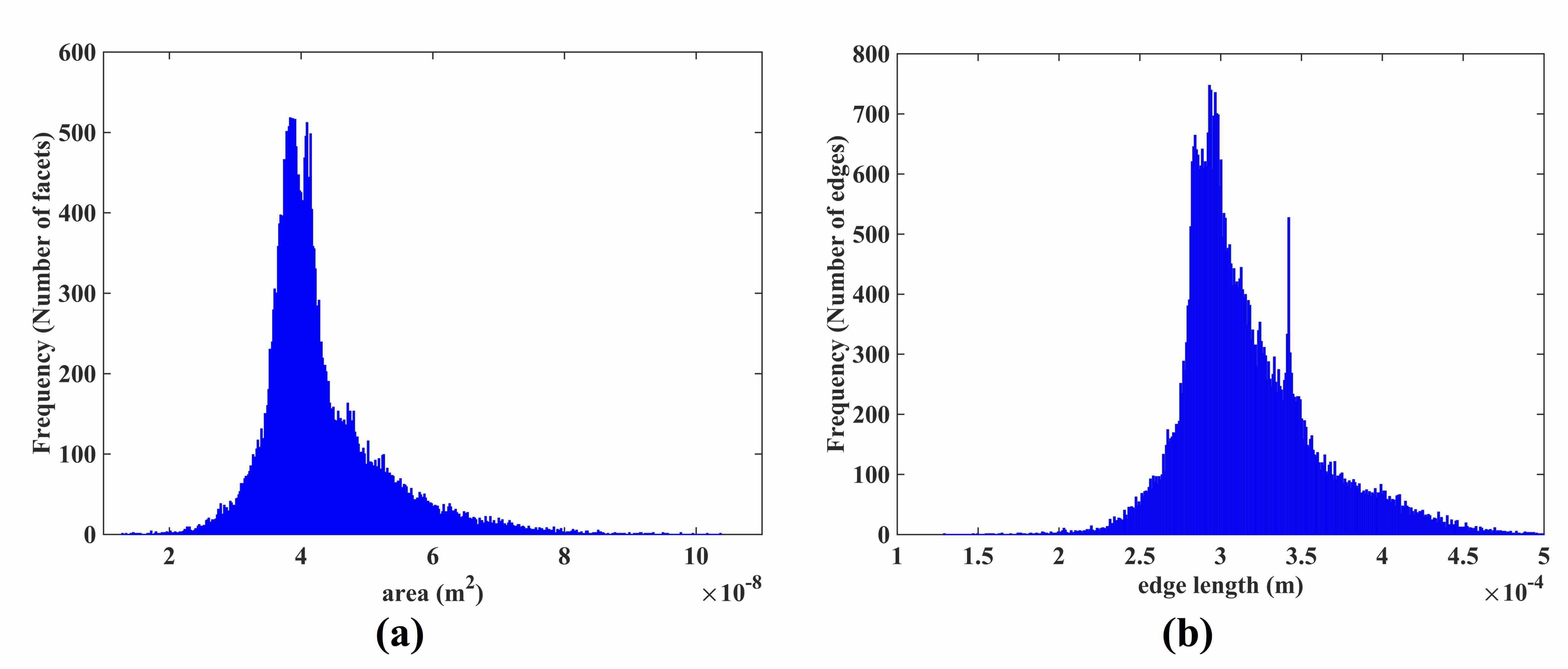}
    \caption{
        U-bend flow problem (Fig.5). Histogram plot for the
        \textbf{(a)}
        edge length and
        \textbf{(b)} 
        area of the immersed boundary facets.
    }
    \label{fig:histogram}
\end{figure}

The tetrahedral mesh is locally refined in the vicinity and within the interior of the
immersed boundary, using the FEniCS built-in function $\it{Refine}$ \cite{unknown_2019_fenics} which uses the algorithm by Plaza and Carey \cite{PLAZA2000195}. The local refinement (tetrahedral mesh resolution
\(h\) close to the Lagrangian points) is based on the surface area
\(\Delta S\) of each Lagrangian point, such that
\(h = a\sqrt{\Delta S}\), with \(0.1 \leq a \leq 2.5\) (see Section 2.5).
We uniformly distribute the Lagrangian points, and we use as
\(\Delta S\) the mean value of the surface area of the facets of the immersed boundary. We examine the distribution
of the Lagrangian points through a histogram plot of the area of the
facets of the immersed boundary. Figure~\ref{fig:histogram}a shows the
histogram plot for the edge length of the immersed boundary facets,
while Fig.~\ref{fig:histogram}b shows the histogram plot for the area of the immersed
boundary facets for the finest Lagrangian point cloud (16,760 nodes and
31,462 facets). For this point cloud, the mean value for the area of the facets
is \({\Delta S}_{\text{mean}} = 4.2848058 \times 10^{- 8}\) and
\({\Delta S}_{\text{std}} = 8.96531 \times 10^{- 9}\). The tetrahedral mesh
resolution \(h\) in the vicinity and inside the immersed boundary that corresponds to the
\(\sqrt{{\Delta S}_{\text{mean}}} = 2.06997 \times 10^{- 4} \approx 2 \times 10^{- 4}\)
(see Table~\ref{tab:ubend-grids}) is $h=2 \times 10^{4}$, such that ratio of
\(\frac{h}{\sqrt{\Delta S_{\text{mean}}}} \approx 1\).

\begin{table}
    \centering
    \caption{
        U-bend flow probblem (Fig.5). Tetrahedral mesh resolution, 
        number of Eulerian nodes, 
        number of tetrahedral linear elements, and
        number of Lagrangian points 
        for the successively denser grids.
    }
    \begin{tabularx}{\textwidth}{@{}XYYYY@{}}
        \toprule
        $h$ (m) & Eulerian & Tetrahedral & Lagrangian & Surface  \\
              & nodes   & elements      & points     & facets \\
        \midrule
        $4\times10^{-4}$ & 10,227 & 59,275 & 4,410 & 8,938 \\
        $2\times10^{-4}$ & 71,589 & 414,928 & 16,760 & 33,516 \\
        $1\times10^{-4}$ & 456,505 & 2,693,729 & 63,082 & 125,848 \\
        \bottomrule
    \end{tabularx}
    \label{tab:ubend-grids}
\end{table}

We use successively denser tetrahedral meshes to obtain a mesh independent numerical solution. 
The meshes used are listed in Table \ref{tab:ubend-grids}.
We use a parabolic velocity profile
(\(u_z(r) = U_{\max}\left(1 - \left( \frac{r}{R} \right)^{2} \right)\), 
\(U_{\max} = 0.122625~\mathrm{m/s}\))
at the inlet (red circle in Fig.~5a), while at the outlet (green circle
in Fig.~5a) we apply zero pressure \(\left( p = 0 \right)\). At the
remaining boundaries (shown in yellow in Fig.~5a) of the flow
domain, we set no-slip boundary conditions. The
(rectangular) flow domain has rigid walls (yellow faces in Fig.~5a)
with no-slip boundary conditions, an inlet (red circle in Fig.~5a),
where the fluid enters, and an outlet (green circle in Fig.~5a),
where the fluid exits. The numerical solutions obtained using the
the three meshes described in Table 3, were projected/interpolated on a body-fitted mesh consisting of 808,984 linear tetrahedral elements and 143,406 nodes.

Fig.~7a shows the axial velocity computed with the proposed method 
using successively denser meshes listed in Table~\ref{tab:ubend-grids}.
Fig.~7b shows a comparison of the axial velocity computed using the IB method, and the experimental data (red dots) at the tube outlet.
The axial velocity is computed on the center plane at the start of the
outlet extension (see Fig.~4a), where in excellent agreement between the velocity predicted using the proposed IB method and experimental results is observed. The numerical results obtained
using the linear (\(P_{1}/P_{1}\)) elements are also in good agreement with the
experimental data, highlighting the accuracy of the proposed method.

\begin{figure}
    \centering
    \includegraphics[width=\textwidth]{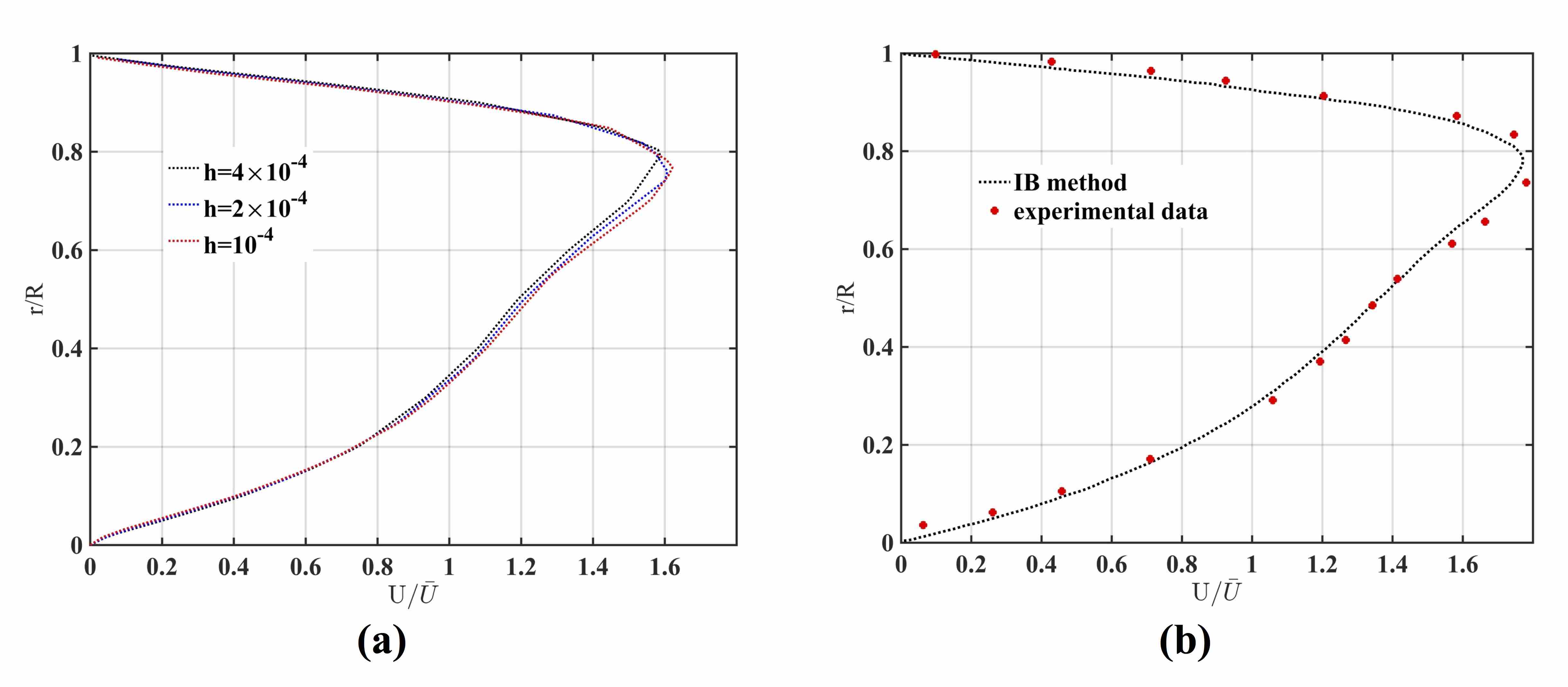}
    \caption{
        \textbf{(a)}
        Axial velocity computed with the proposed IB method
        using successively denser meshes described in Table~\ref{tab:ubend-grids} with $h$ being the mesh resolution close to the immersed boundary, and 
        \textbf{(b)}
        axial velocity computed with the proposed IB method (dashed line) 
        compared against the experimental data (red dots) at the tube outlet for the U-bend flow problem.
    }
    \label{fig:my_label}
\end{figure}

\section{Numerical examples: blood flow simulations}

In this section, we demonstrate the accuracy and robustness of the
proposed scheme through blood flow simulations in complex vascular
geometries. In the first flow problem, we simulate blood flow in
coronary artery bifurcation (Fig. 8), while in the second we consider blood flow
in the ascending aorta (Fig. 11). Finally, we consider the flow in the ascending
and descending aorta (Fig. 14).

The flow problems considered demonstrate the applicability of the
proposed method for internal flow simulations. Internal flow cases, are
practically out of reach for all IB methods. The computational cost of
having a tetrahedral mesh (uniform or locally refined) to discretize the
flow domain is high, since only a small percentage of the total
number of mesh elements, falls into the domain of interest
\citep{zhu_etal_2019_graphpartitioned}.
The
proposed IB method allows for cropping the tetrahedral mesh (see example 4.3) that covers the complex (vascular) geometry, and at the same time increases the percentage of elements that are located inside the flow domain of interest. In all
flow cases considered, we use a locally refined mesh, in the vicinity and inside
the Lagrangian points. We demonstrate the accuracy of the proposed IB
method, by comparing the numerical findings with those computed using a
body-fitted mesh FE flow solver. Local refinement is an automated
procedure implemented using FEniCS built-in functions (Plaza algorithm \cite{PLAZA2000195}).

\subsection{Flow in an coronary artery bifurcation}

In the first example, we simulate the blood flow in a coronary
artery (CA) bifurcation (shown in Fig.~8). The immersed boundary
(coronary artery) is embedded into a box with dimensions
\(- 0.0052 \ m \leq x \leq 0.01 \ m\), \(- 0.006 \ m \leq y \leq 0.006 \ m\) and
\(0 \ m \leq z \leq 0.0296 \ m\). The inlet and
two outlets of the CA geometry were extruded to ensure that the flow is
fully developed.

\begin{table}
    \centering
    \caption{
        Mesh resolution close to the immersed object, 
        number of tetrahedral linear elements,
        mesh points, and
        number of Lagrangian points 
        for the successively denser grids
        considered for the flow in the coronary artery bifurcation. 
    }
    \begin{tabularx}{\textwidth}{@{}XYYYY@{}}
        \toprule
        $h$ (m) & Eulerian & Tetrahedral & Surface  & Lagrangian \\
              & nodes   & elements    & facets   & points \\
        \midrule
        $4\times10^{-4}$ & 18,364 & 117,864 & 3,689 & 1,770 \\
        $2\times10^{-4}$ & 108,897 & 627,040 & 7,822 & 3,913 \\
        $1\times10^{-4}$ & 618,565 & 3,641,633 & 14,894 & 7,449 \\
        \bottomrule
    \end{tabularx}
    \label{tab:coronary-grids}
\end{table}

We discretize the flow domain using tetrahedral elements. The box
domain is discretized with a high quality tetrahedral mesh, generated using a uniform Cartesian grid (to generate the mesh we used the FEniCS build-in function BoxMesh). The tetrahedral mesh is locally refined in the vicinity and inside the immersed boundary (Figs. 8c-d). We use
successively denser meshes to obtain a mesh independent numerical
solution. The meshes used are listed in Table \ref{tab:coronary-grids}.
The numerical results on the two finest meshes (for 627,040 and 3,641,633) are indistinguishable.
Mesh generation is a straightforward and automated.
The locally refined mesh with 3,641,633 elements
is generated in less than 60 s (using a laptop computer with 2.7 GHz i7 quad-core processor and 16 GB internal memory),
using the refine function in FEniCS. 
We define the mesh resolution close to the immersed object and in the interior
based on the mean value of the surface elements area (see Section 3.2).
The majority (85\%) of the elements are located inside the immersed boundary (coronary artery).

\begin{figure}
    \centering
    \includegraphics[width=\textwidth]{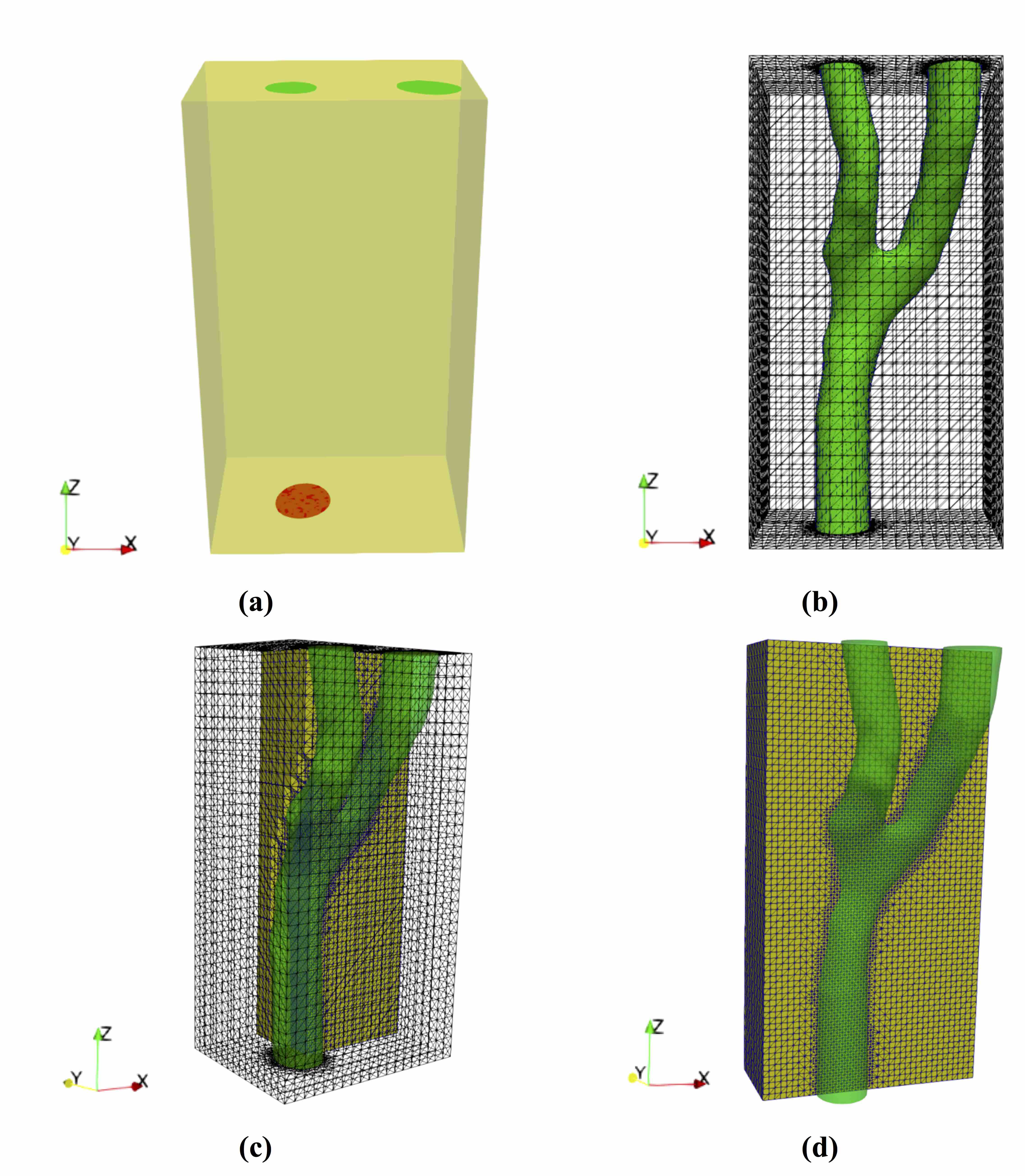}
    \caption{
    Flow domain for the coronary artery bifurcation example, with 
        inlet (circle in red color), 
        outlet (circle in green color) and
        walls (yellow color) boundaries; 
        \textbf{(b)}
        wireframe of the tetrahedral mesh 
        and the coronary artery immersed boundary (Lagrangian points);
        \textbf{(c)}
        tetrahedral mesh refined close and inside the immersed boundary, and
        \textbf{(d)}
        vertical cross-section of the tetrahedral mesh.}
    \label{fig:my_label}
\end{figure}

We set the total time for the simulation to \(T = 0.8\) s (one
cardiac cycle), and the time step to \(dt = 5 \times 10^{- 4}\) s. At the
inlet ((red area located at the bottom surface in Fig.(8a)) we apply the pulsatile
velocity waveform shown in Fig.~9 (we use a parabolic velocity profile
since the Womersley number is small), and zero pressure boundary
conditions at the two outlets (green area on the top surface in Fig.(8a)). At the
remaining walls (yellow surface in Fig.(8a)), we apply no-slip boundary conditions.
We employ the Newtonian model for blood flow, with dynamic viscosity of
\(\mu = 0.00345\ Pa \cdot s\) and density of
\(\rho = 1,056\ \frac{\text{kg}}{m^{3}}\).

\begin{figure}
    \centering
    \includegraphics[width=\textwidth]{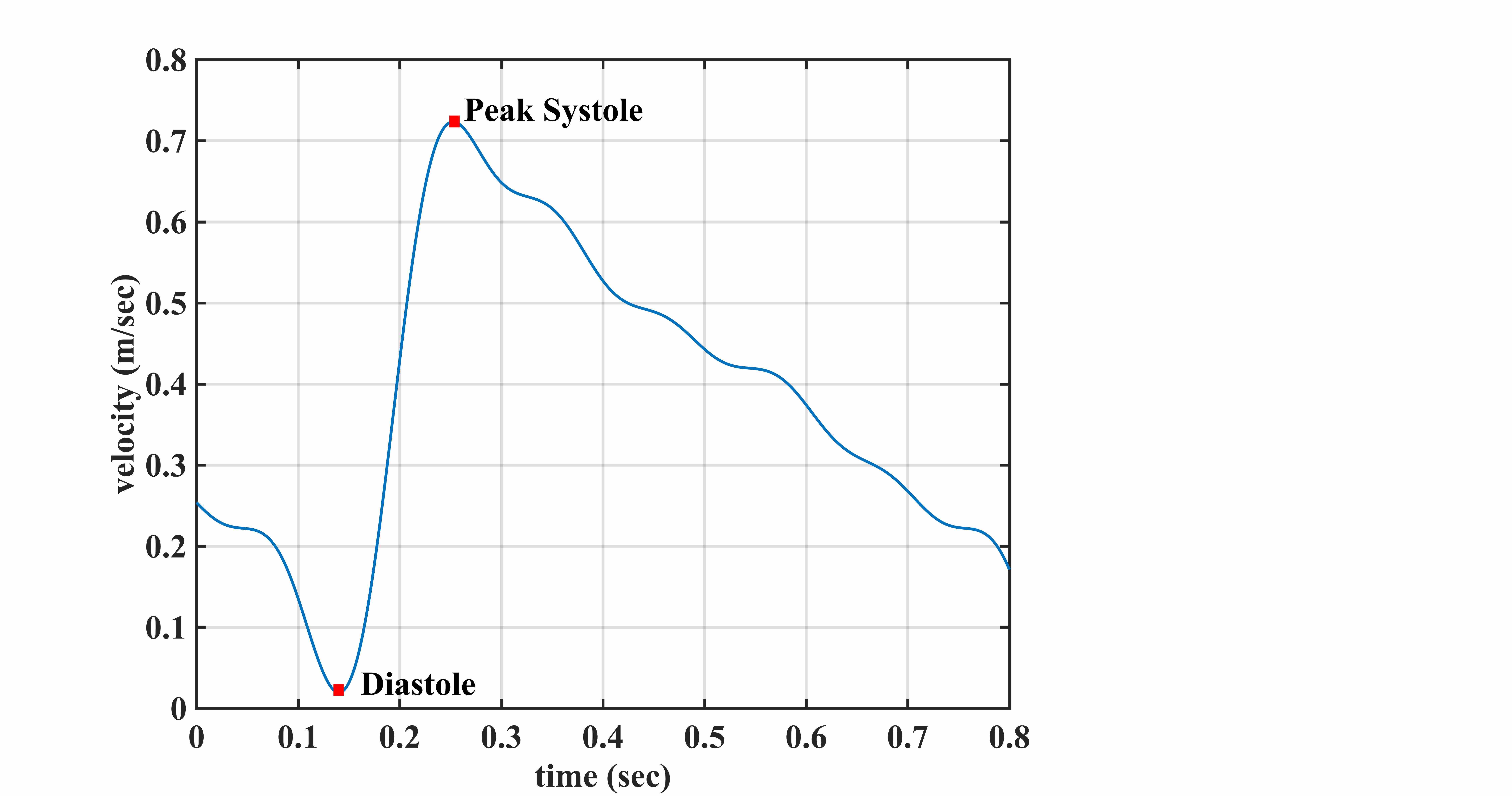}
    \caption{
        Pulsatile velocity waveform imposed at the inlet for the coronary artery bifurcation flow example.
    }
    \label{fig:my_label}
\end{figure}

To demonstrate the accuracy of our IB method, we compare the
results obtained using this method with those for the body-fitted mesh. The
body-fitted mesh consists of 154,681 nodes and 860,818 linear
tetrahedral elements \(\left( P_{1}/P_{1} \right)\). Table~\ref{tab:coronary-norms}, lists the
normalized root mean square error (NRMSE)
\(L_{\text{NRMSE}} = \frac{1}{N}\sqrt{\sum_{i = 1}^{N}\left( u_{i}^{\text{IB}} - u_{i}^{\text{body\ fitted}} \right)^{2}}\)
error norms at different time instances.

\begin{table}
    \centering
    \caption{\(L_{2}\) error norm
        of the immersed boundary method numerical results 
        against those computed using the body-fitted FE method
        at different time instances for the coronary artery bifurcation flow example.
    }
    \begin{tabular*}{\textwidth}
    {@{\extracolsep{\fill}} ccccccccc }
        \toprule
        $L_2$ (\%) & $t=0.1$ & $t=0.2$ & $t=0.3$ & $t=0.4$ & $t=0.5$ & $t=0.6$ & $t=0.7$ & $t=0.8$ \\
        \midrule
        $u_x$ & 1.53 & 1.38 & 2.03 & 1.94 & 1.94 & 1.88 & 1.84 & 1.74 \\
        $u_y$ & 2.32 & 3.17 & 3.34 & 3.27 & 3.28 & 3.23 & 3.16 & 2.78 \\
        $u_z$ & 3.34 & 3.82 & 4.57 & 4.42 & 4.44 & 4.37 & 4.06 & 3.84 \\
        \bottomrule
    \end{tabular*}
    \label{tab:coronary-norms}
\end{table}

The results obtained using our IB method and body fitted mesh reported in Table 5 are for practical purposes indistinguishable as in the bioengineering applications. The differences of under 5$\%$ would be overwhelmed by the biomechanical properties uncertainties. The decisive advantage of our method is the easiness of patient-specific computational grid generation with practically no increase in the computational cost. The only computational overhead  of our IB method in comparison to the approach relying on a body fitted mesh is due to the numerical solution of the linear system given by Eq.(21). For the computation of flow in coronary artery bifurcation  conducted here (Figure 8), this overhead was negligible — less than 1$\%$ of the total computation time. Fig.~10 shows the streamlines in the coronary artery bifurcation at time \(t = 0.14\)
(diastole) and \(0.25\) sec (peak systole).

\begin{figure}
    \centering
    \includegraphics[width=\textwidth]{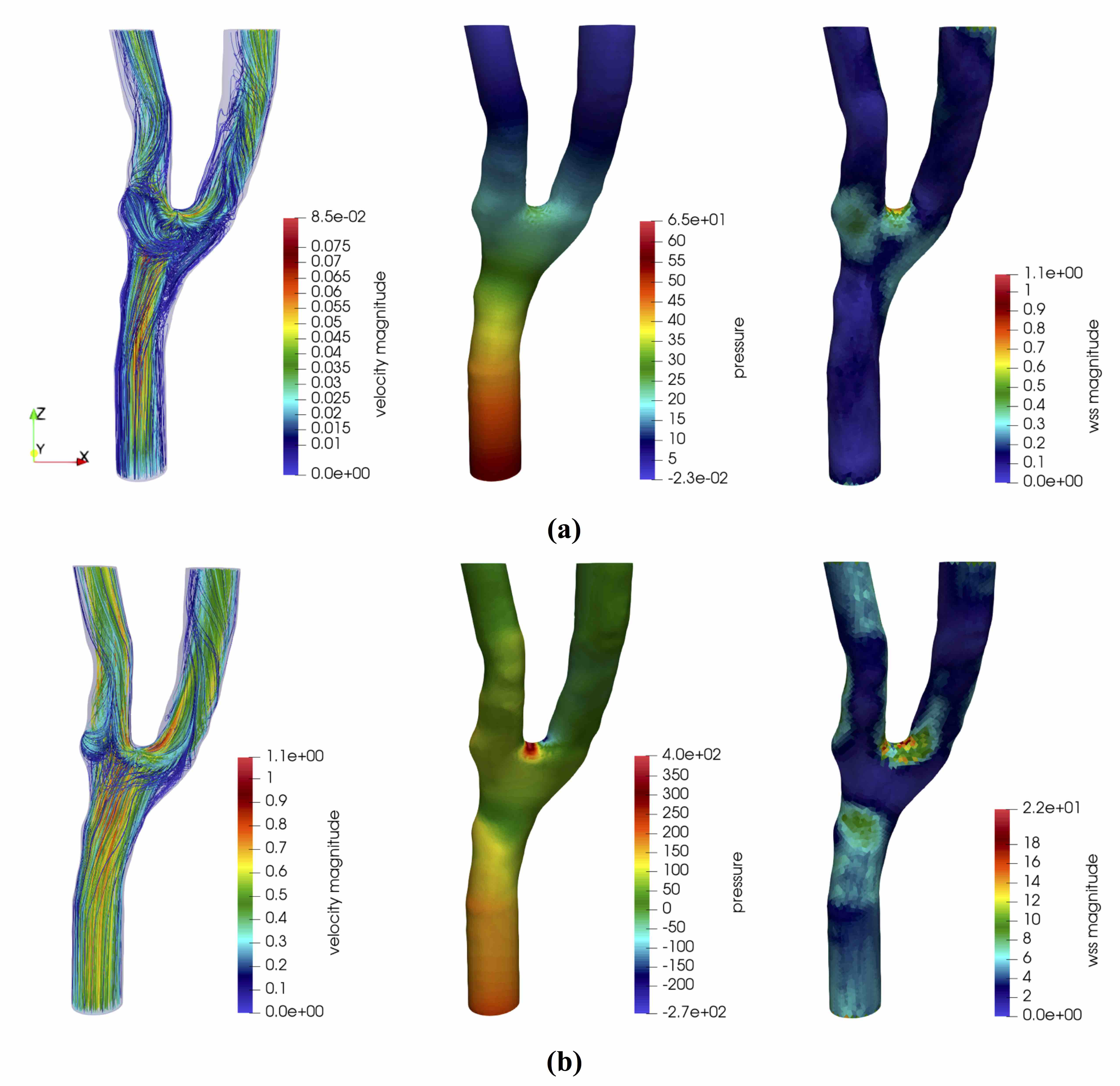}
    \caption{
        Streamlines (left), 
        pressure (middle) and
        wall shear stress (WSS) magnitude (Pa) (right) at
        \textbf{(a)}
        peak systole pressure and
        \textbf{(b)}
        diastolic minimum for the coronary artery bifurcation flow example.
    }
    \label{fig:my_label}
\end{figure}

\subsection{Blood flow in the ascending aorta}

In the second example, we simulate blood flow in the ascending aorta
(the brachiocephalic, left common carotid, and left subclavian arteries are included in the model) shown in Fig.~11. The immersed boundary (ascending aorta) is
embedded within a box (Fig.~11a) with dimensions
\(- 0.037 \ m \leq x \leq 0.035 \ m\), \(- 0.08 \ m \leq y \leq 0.04 \ m\) and
\(0 \ m \leq z \leq 0.12 \ m\). The box domain is discretized with a high quality tetrahedral mesh, generated using a uniform Cartesian grid (to generate the mesh we used the FEniCS build-in function BoxMesh). The tetrahedral mesh is locally refined in the vicinity and inside the immersed boundary as shown in Fig.11c-d. We  use  successively  denser meshes, indicated in Table 6, to obtain a mesh independent numerical solution. The numerical results on the two finest meshes (for 1,410,846 and 2,557,306) are indistinguishable. We refine the tetrahedral mesh close to the Lagrangian points and in the
interior of the immersed boundary. As described in Section 2.5, the mesh
resolution close to the immersed boundary (Lagrangian points) is based
on on the mean area of the facets on the immersed object. Mesh generation and local refinement on the rectangular domain is a straightforward and efficient procedure. It took less than 30 s to generate the finest grid (information about this grid in the third row of Table 6) using a laptop computer with i7 quad-core processor with 16 GB internal memory.

\begin{figure}
    \centering
    \includegraphics[width=\textwidth]{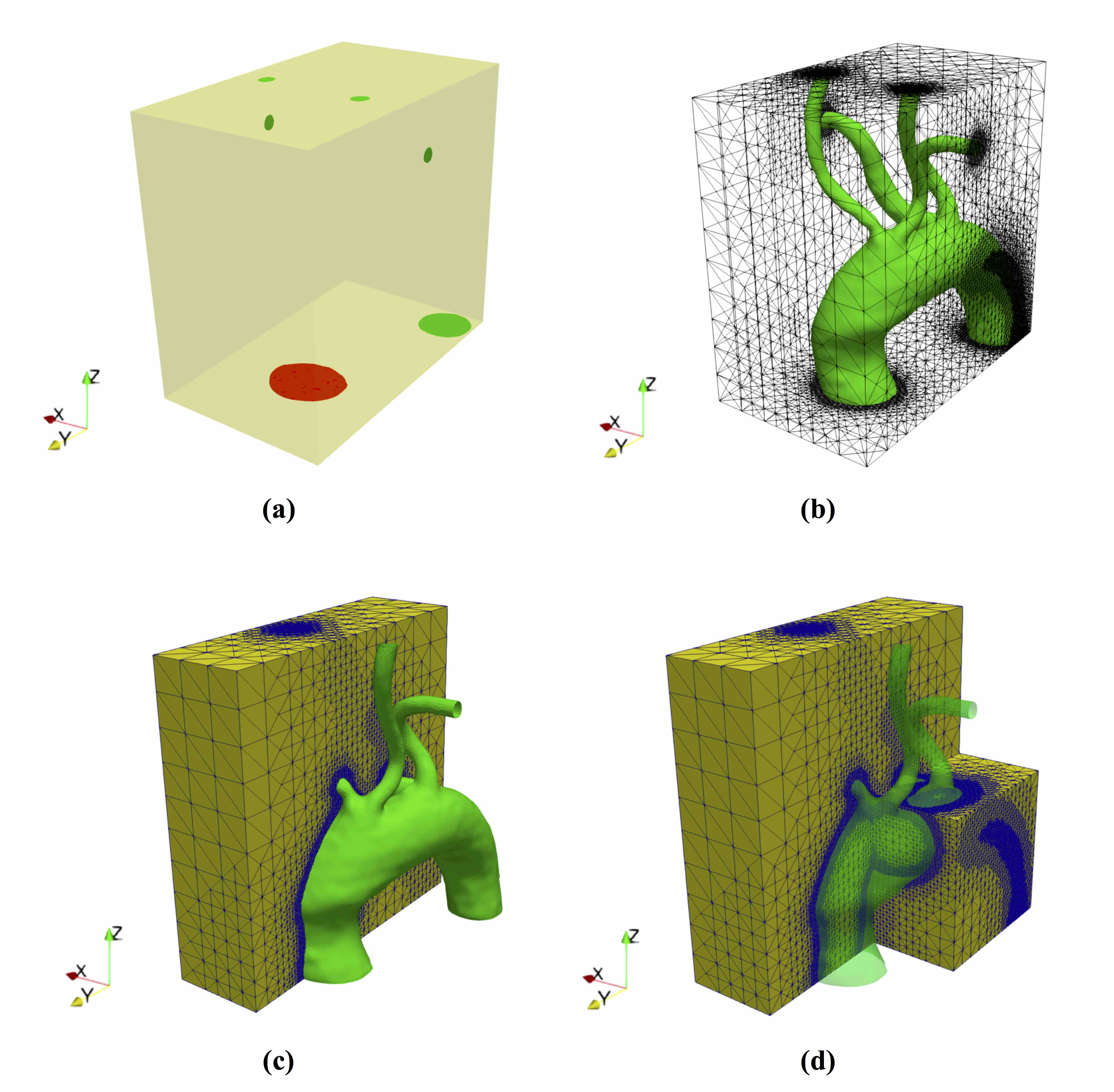}
    \caption{
        \textbf{(a)}
        Flow domain for the ascending aorta flow example, with
        inlet (circle in red color), 
        outlets (circle in green color) and
        walls (yellow color) boundaries;
        \textbf{(b)} 
        wireframe of the tetrahedral mesh and
        the immersed boundary (ascending aorta);
        \textbf{(c)}
        horizontal cross-section of the tetrahedral mesh, and
        \textbf{(d)}
        vertical and horizontal cross-sections of the tetrahedral mesh.
    }
    \label{fig:my_label}
\end{figure}

\begin{table}
    \centering
    \caption{
        Mesh resolution close to the immersed object, 
        number of tetrahedral mesh vertices, and 
        number of Lagrangian points for the successively denser grids 
        considered for the flow in the ascending aorta.
    }
    \begin{tabularx}{\textwidth}{@{}XYYYY@{}}
        \toprule
        $h$ (m) & Eulerian & Tetrahedral & Surface  & Lagrangian \\
              & nodes   & elements    & facets   & points \\
        \midrule
        $1.5 \times10^{-3}$ &  61,930 &   355,296 &  7,410 &  3,584 \\
        $7.5 \times10^{-4}$ & 241,483 & 1,410,846 & 15,153 &  7,641 \\
        $3.75\times10^{-4}$ & 439,126 & 2,557,306 & 31,684 & 15,938 \\
        \bottomrule
    \end{tabularx}
    \label{tab:aorta-grids}
\end{table}

We set the
total time for the simulation to \(T = 0.8\)~s (one cardiac
cycle), and the time step \(dt = 5 \times 10^{- 4}\)~s. At the inlet
(red area in Fig.~11a) we apply the pressure waveform shown in Fig.~12,
zero pressure boundary conditions at the two outlets (green area in Fig.~11a), and no-slip boundary conditions at the remaining walls (yellow
surface in Fig.~11a). We employ the Newtonian model for blood flow, with
dynamic viscosity of \(\mu = 0.00345\ \mathrm{Pa \, s}\) and density of
\(\rho = 1,056\ \mathrm{kg/m^3}\). We use the diameter \(D\) and
the maximum velocity \(U_{m}\) as the characteristic length and velocity
to calculate the Reynolds number.

\begin{figure}
    \centering
    \includegraphics[width=\textwidth]{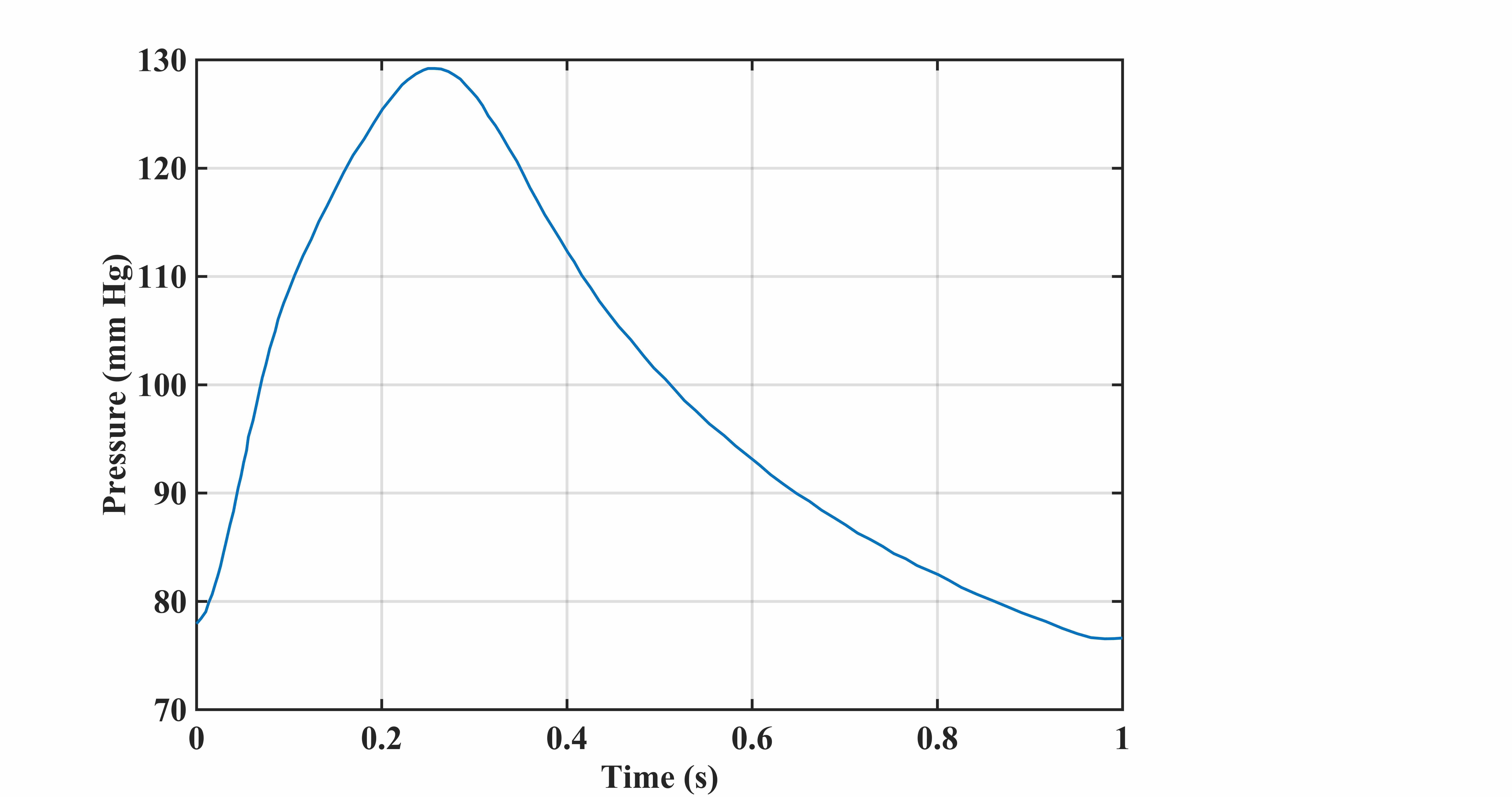}
    \caption{
        Pressure waveform used in the ascending aorta flow example.
    }
    \label{fig:my_label}
\end{figure}

To compare the flow velocity predicted using our IB method with the results from body fitted mesh, we interpolate our IB solution on a body fitted mesh consisting of 92,786 nodes and 504,628 linear tetrahedral elements. The comparison done at the time instances (every 0.1 s interval) indicated that the normalized root mean square error (NRMSE) of less than 5$\%$ for the three velocity components. We conducted also qualitative comparison of the predicted streamlines, pressure contours, and wall shear stress (WSS) contours at the time instances of \(t = 0.2\) and \(t = 0.6\) (Fig. 13). No visually distinguishable differences between the results obtained using our IB method and approach using body fitted mesh can be found in Fig. 13. (every 0.1 sec) the IB numerical solution with the numerical results obtained using the body-fitted mesh. The normalized root mean
square error (NRMSE) \(L_{\text{NRMSE}}\) for the three velocity
components is less than 5\%.

\begin{figure}
    \centering
    \includegraphics[width=\textwidth]{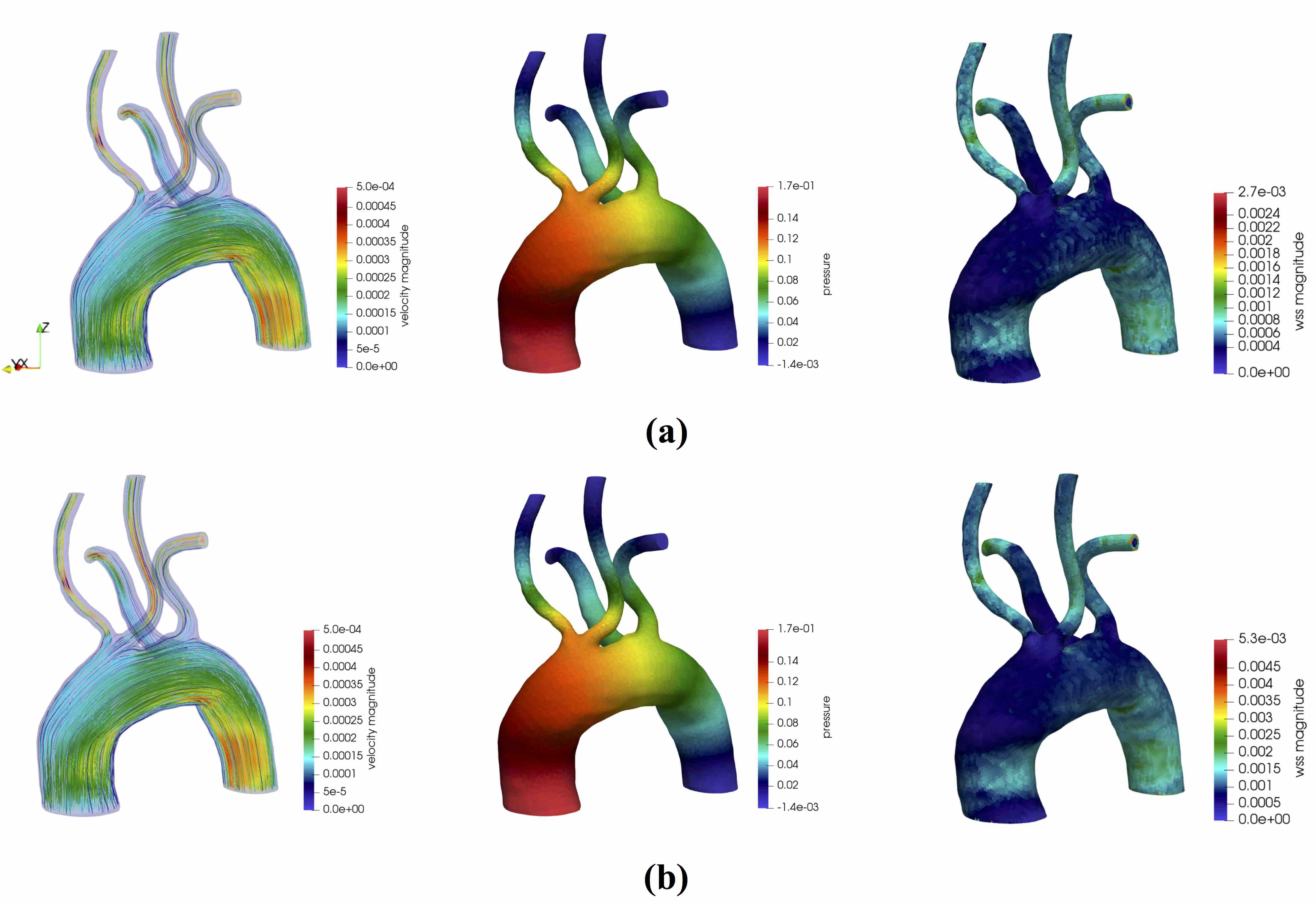}
    \caption{
        Streamlines (left), 
        pressure (middle) and
        wall shear stress (WSS) magnitude (Pa) (right) 
        contour plot at time
        \textbf{(a)}
        \(t = 0.2\)~s and
        \textbf{(b)}
        \(t = 0.6\)~s for the ascending aorta flow example.
    }
    \label{fig:my_label}
\end{figure}

Fig.~13 shows the streamlines (left), pressure (middle) and wall shear
stress (WSS) magnitude (right) contours at different time instances
\(t = 0.2\) and \(t = 0.6\). The results were obtained using the mesh with
1,410,846 tetrahedral elements.

\subsection{Blood flow in the ascending and descending aorta}

As the final example, we consider blood flow in the ascending and
descending aorta. This example differs from the previous one (blood flow in
the ascending aorta) by the number of outlets considered (the
brachiocephalic, left common carotid, and left subclavian arteries were
removed).

This flow example is more challenging because the inlet of the immersed
object is located in the interior of the box flow domain and not
on the boundary (Fig.~14a). Therefore, we crop the rectangular domain
such that the inlet of the immersed object (red circle in Fig.~14a) is
on the boundary of the flow domain (yellow color in Fig.~14a) and the
boundary conditions can be easily applied.

\begin{figure}
    \centering
    \includegraphics[width=\textwidth]{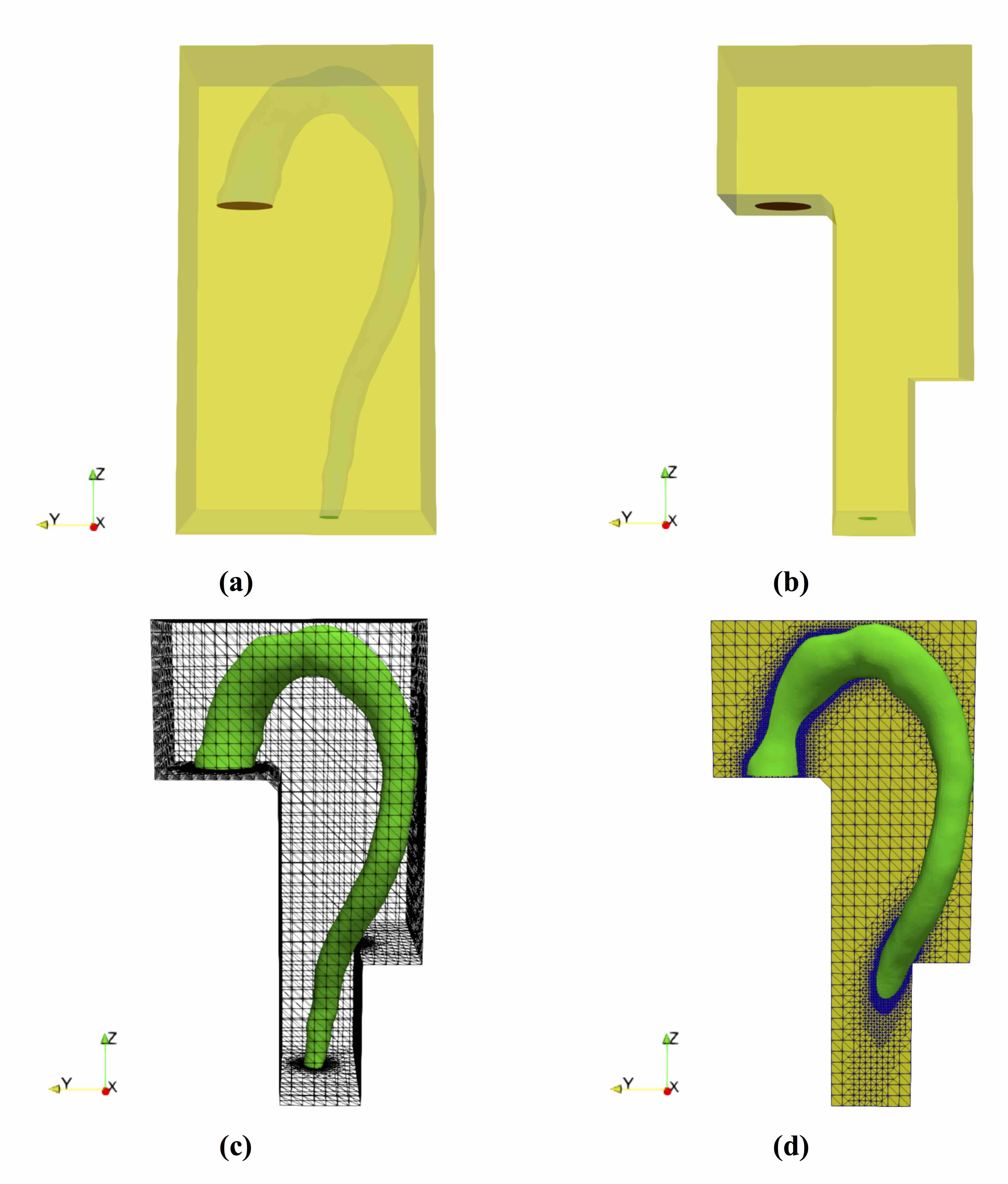}
    \caption{
        \textbf{(a)} 
        Flow domain
        for the ascending and descending aorta flow example, with
        inlet (circle in red color),
        outlets (circle in green color) and
        walls (green color) boundaries;
        \textbf{(b)}
        cropped flow domain; 
        \textbf{(c)} 
        wireframe of the tetrahedral mesh and
        the immersed boundary
        (ascending and descending aorta) and,
        \textbf{(d)}
        horizontal cross-section of the tetrahedral mesh.
    }
    \label{fig:my_label}
\end{figure}

The immersed boundary (aorta) is embedded into a box domain
(Fig.~14a) with dimensions \(- 0.042 \ m \leq x \leq 0.0252 \ m\),
\(- 0.0806 \ m \leq y \leq 0.025 \ m\) and \(- 0.14 \ m \leq z \leq 0.0616 \ m\). The box domain is discretized with a high quality tetrahedral mesh, generated using a uniform Cartesian grid (to generate the mesh we used the FEniCS build-in function BoxMesh). The tetrahedral mesh is cropped and elements are removed (the cropped domain is shown in Fig.~14b). Finally,
the cropped tetrahedral mesh is locally refined in the vicinity and inside the immersed boundary (Fig.~14d).

\begin{figure}
    \centering
    \includegraphics[width=\textwidth]{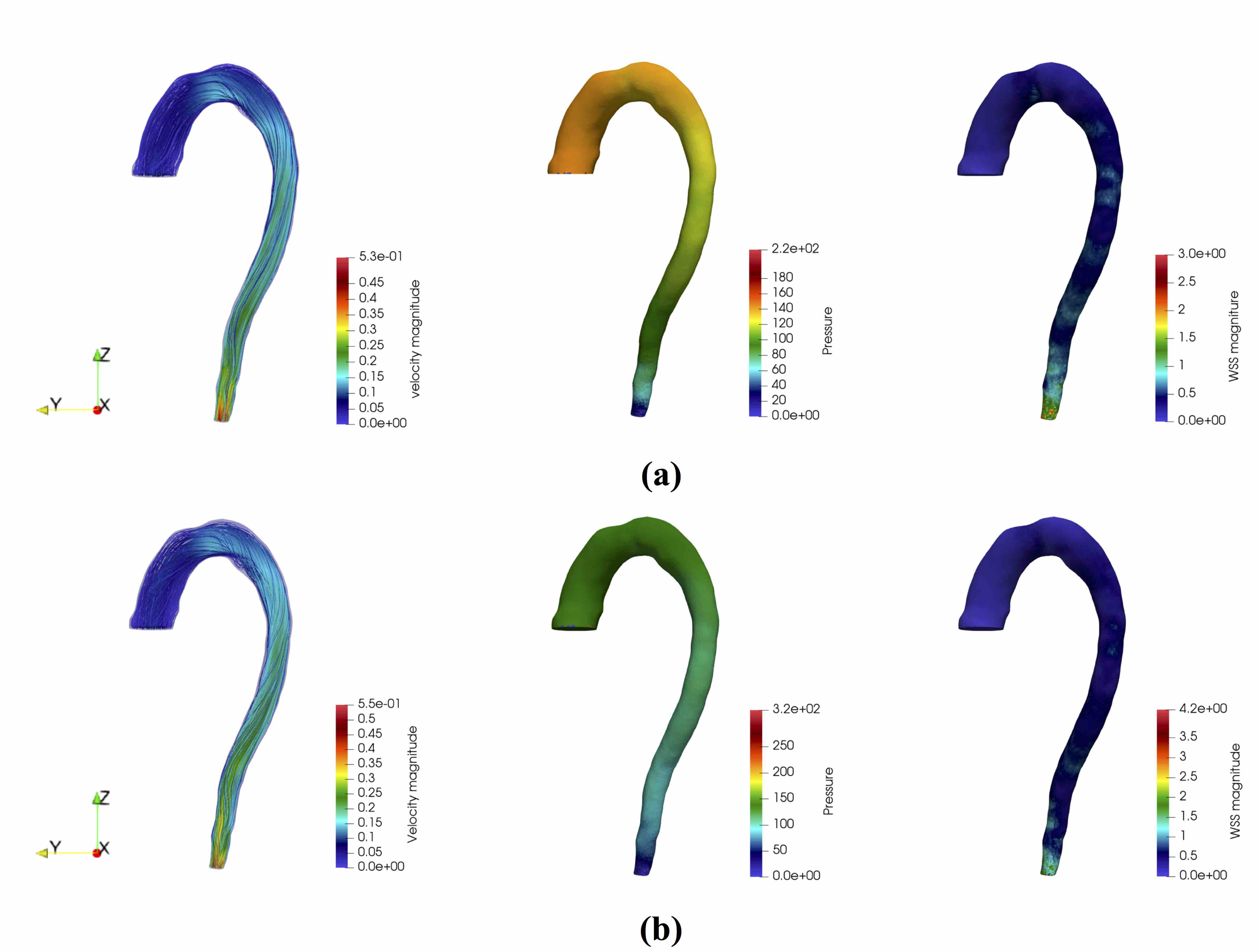}
    \caption{
        Streamlines (left),
        pressure (middle) and 
        wall shear stress (WSS) magnitude (Pa) (right) 
        contour plot at time
        \textbf{(a)}
        \(t = 0.2\) s and
        \textbf{(b)}
        \(t = 0.6\) s for the ascending and descending aorta flow example.
    }
    \label{fig:my_label}
\end{figure}

We use successively denser meshes to obtain a mesh independent numerical
solution. A mesh of 417,227 nodes and 2,409,460 linear tetrahedral
elements (approx.~85\% of the elements are located inside the
immersed boundary) ensures a mesh independent numerical solution. We
apply a uniform velocity profile \(\overline{U}\) (plug flow) at the
inlet (in red color Fig.~14a). We apply zero pressure boundary
conditions at the outlet, and no-slip boundary conditions in the
remaining boundaries. We employ the Newtonian model for blood flow, with
dynamic viscosity of \(\mu = 0.00345\ \mathrm{Pa \cdot s}\) and density of
\(\rho = 1,056\ \mathrm{kg/m^3}\). The diameter \(D\) at the
inlet, and the velocity at the inlet \(\overline{U}\), are used as the
characteristic length and velocity to define the Reynolds number. For
the inlet velocity value used in our simulations, the Reynolds number is
\(\mathrm{Re} = 310\) for the present geometry.

Fig.~15 shows the streamlines (left), pressure (middle) and wall shear
stress (WSS) magnitude (right) contours at different time instances
\(t = 0.2\) and \(t = 0.6\).

\section{Conclusions}

In this work, we present an immersed boundary method for internal
flows, showing blood flow application. The proposed scheme combines the finite element (FE) method,
which is used to numerically solve the Navier--Stokes equations, and the
boundary condition enforced immersed boundary (BCE-IB) method to account for
complex geometries.

The proposed method increases the computational efficiency of the IB
method for internal flows, by reducing the number of unused mesh points
through a sophisticated and efficient method for mesh
refinement in the vicinity and inside the immersed boundary. This method facilitated generation of a computational mesh consisting of over three and half million elements (computation of flow in an aortic bifurcation — see Fig. 8) in less than 30 seconds on an off-the-shelf laptop with quad-core i7 processor and 16 GB of internal memory. This very promising results make it possible to regard our adaptive mesh refinement method as compatible with the time constraints of clinical workflow (no firm definition here by the times of on order of under 1 minute have been reported in the literature \cite{WITTEK2010292}). Furthermore,
we introduced an efficient method to reduce (crop) the flow domain, in flow
cases where the immersed boundary inlet(s) and outlet(s) do not coincide
with the boundaries of the flow domain (tetrahedral mesh).

The proposed method utilizes an efficient and accurate finite element
solver based on incremental pressure correction scheme (IPCS). All simulations in this study were conducted using an off-the-shelf laptop with quad-core i7 processor and 16 GB of internal memory. The
accuracy of the proposed scheme has been successfully verified through comparison
with an analytical solution for the Poiseuille flow in a cylinder with
circular cross section. We further validated the accuracy of the proposed
method by solving a flow case where experimental data are available.
The applicability and efficiency of the proposed method has been
demonstrated through the solution of flow examples with complex geometry.

\clearpage
\section*{Acknowledgments}

A. Wittek and K. Miller acknowledge the support by the Australian Government through the
Australian Research Council's Discovery Projects funding scheme (project
DP160100714). The views expressed herein are those of the authors and
are not necessarily those of the Australian Research Council.

\bibliography{references}



\end{document}